\documentclass[smallextended]{svjour3}

\usepackage{verbatim}
\usepackage[table]{xcolor}
\usepackage{listings}
\usepackage{amssymb}
\usepackage{caption}
\usepackage{balance}
\usepackage{tabularx}
\usepackage{booktabs}
\usepackage{color}
\usepackage{colortbl}
\usepackage{mathtools}
\usepackage{multirow}
\usepackage{rotating}
\usepackage{graphicx}
\usepackage{url}
\usepackage{tipa}
\usepackage[framemethod=tikz]{mdframed}
\usepackage{balance}

\usepackage[numbers]{natbib}

\newcommand{\ie}{i.e.,\space}
\newcommand{\eg}{e.g.,\space}

\newcommand{\etal}{et al.\space}

\definecolor{lightgray}{gray}{0.9}

{

	\newmdenv[innerlinewidth=0.5pt, roundcorner=4pt,innerleftmargin=6pt,
	innerrightmargin=6pt,innertopmargin=6pt,innerbottommargin=6pt]{mybox}
	
	\definecolor{pgreen}{rgb}{0,0.5,0}
	
	\lstset{language=Java,
		basicstyle=\scriptsize\ttfamily,
		keywordstyle=\color{blue},
		commentstyle=\color{pgreen}\ttfamily,
		stringstyle=\ttfamily\color{red!50!brown},
		numbers=left,
		numberstyle=\tiny\color{gray},
		stepnumber=1,
		numbersep=8pt,
		showstringspaces=false,
		breaklines=true,
		frameround=ftff,
		frame=single,
		belowcaptionskip=.25\baselineskip}
	
\definecolor{gray50}{gray}{.5}
\definecolor{gray40}{gray}{.6}
\definecolor{gray30}{gray}{.7}
\definecolor{gray20}{gray}{.8}
\definecolor{gray10}{gray}{.9}
\definecolor{gray05}{gray}{.95}
	
\sloppy

\begin{document}

\title{Improving Change Prediction Models with Code Smell-Related Information}
\titlerunning{Improving Change Prediction Models with Code Smell-Related Information}

\author{Gemma Catolino \and Fabio Palomba \and Francesca Arcelli Fontana \and Andrea De Lucia \and Andy Zaidman \and Filomena Ferrucci}

\institute{
	Gemma Catolino, Andrea De Lucia, Filomena Ferrucci \at
	University of Salerno, Italy\\
	\email{gcatolino@unisa.it, adelucia@unisa.it, fferrucci@unisa.it} 
	\and
	Fabio Palomba \at
	University of Zurich, Switzerland\\
	\email{palomba@ifi.uzh.ch}
	\and
	Francesca Arcelli Fontana \at
	University of Milano-Bicocca, Italy\\
	\email{arcelli@disco.unimib.it}
	\and
	Andy Zaidman \at
	Delft University of Technology, The Netherlands\\
	\email{a.e.zaidman@tudelft.nl}
}

\date{Received: date / Accepted: date}
\maketitle

\begin{abstract}
Code smells represent sub-optimal implementation choices applied by developers when evolving software systems. The nagative impact of code smells has been widely investigated in the past: besides developers' productivity and ability to comprehend source code, researchers empirically showed that the presence of code smells heavily impacts the change-proneness of the affected classes. On the basis of these findings, in this paper we conjecture that code smell-related information can be effectively exploited to improve the performance of change prediction models, \ie models having as goal that of indicating to developers which classes are more likely to change in the future, so that they may apply preventive maintenance actions. Specifically, we exploit the so-called \emph{intensity index}---a previously defined metric that captures the severity of a code smell---and evaluate its contribution when added as additional feature in the context of three state of the art change prediction models based on product, process, and developer-based features. We also compare the performance achieved by the proposed model with the one of an alternative technique that considers the previously defined antipattern metrics, namely a set of indicators computed considering the history of code smells in files. Our results report that (i) the prediction performance of the intensity-including models is statistically better than that of the baselines and (ii) the intensity is a more powerful metric with respect to the alternative smell-related ones. Nevertheless, we observed some complementarities between the set of change-prone and non-change-prone classes correctly classified by the models relying on intensity and antipattern metrics: for this reason, we devise and evaluate a smell-aware combined change prediction model including product, process, developer-based, and smell-related features. We show that this model has an F-Measure that is up to 20\% higher than the existing state-of-the art models.

\end{abstract}

%\newpage

\section{Introduction}
\label{sec:intro}
% !TEX root =  main.tex
During software evolution, change is unavoidable. Indeed, software systems are continuously modified in order to be adapted to changing needs, improved in performance or maintainability, or fixed from potential bugs \cite{lehman:1985}. As a consequence, they become more complex, possibly eroding the original design with a subsequent reduction of their overall maintainability \cite{parnas}. In this context, predicting the source code components having a higher likelihood to change in the future represents an important activity to allow developers to plan preventive maintenance operations such as, \eg refactoring \cite{fowler:1999} or peer-code reviews \cite{bacchelli:2013}. 

For this reason, the research community proposed several approaches in order to allow developers to control these changes \cite{catolino2018enhancing,elish2013suite,eski2011empirical,girba2004yesterday,kumar2017transfer,lu2012ability,malhotra2015predicting,romano2011using,zhou2009examining}. Such approaches are based on the usage of machine learning models exploiting several predictors that capture different characteristics of classes \ie structural, process, and developer-related features. 

Despite the good performance shown by those existing models, recent studies \cite{khomh2012exploratory,palomba2017diffuseness} explored new factors contributing to the change-proneness of classes, finding that it is strongly influenced by the presence of the so-called bad code smells \cite{fowler:1999}, \ie sub-optimal design and/or implementation choices applied by practitioners when developing a software system. Specifically, such studies showed that smelly classes are significantly more likely to be the subject of changes than classes not affected by any design problem. 

In this paper, we capture such findings and empirically investigate the extent to which smell-related information can be actually useful when considered in the context of the prediction of change-prone classes: our conjecture is that \emph{the addition of a measure of code smell severity can improve the performance of existing change prediction models, as it may help in the correct assessment of the change-proneness of classes}. For severity, we mean a metric able to quantify how much a certain code smell instance is harmful for the design of a source code class. To test our conjecture, we (i) add the \emph{intensity index} defined by Arcelli Fontana \etal \cite{fontana2015towards} in three state of the art change prediction models based on structural \cite{zhou2009examining}, process \cite{elish2013suite}, and developer-related metrics \cite{catolino2018enhancing} and (ii) evaluate---on 43 releases of 14 large systems---how much such addition improves the prediction capabilities of the baseline models. 
We then compare the performance of the intensity-including change prediction models with the one achievable by exploiting alternative smell-related information such as the antipattern metrics defined by Taba \etal \cite{Taba:2013}, which are able to capture historical information on code smell instances (\eg the recurrence of a certain instance over time). The results show that the addition of the intensity index provides significant improvements in the performance of the baseline change prediction models, with an increase of 10\% in terms of F-Measure on average. Furthermore, the intensity-including models work better than the models built adding the alternative antipattern metrics, although these two types of information are complementary, \ie they correctly capture the change-proneness of different change-prone smelly classes.

Given such a complementarity, we then further explore the possibility to improve change prediction models by devising a smell-aware combined approach that mixes together the features of the experimented models, \ie structural, process, developer-, and smell-related information, with the aim of boosting the change-proneness prediction abilities. As a result, we discovered that such a combined model is able to improve by up to 20\% the performance of the baseline approaches. 

To sum up, the contributions of this paper are the following:

\begin{enumerate}
	
	\item A large-scale empirical assessment of the role of the intensity index \cite{fontana2015towards} when predicting change-prone classes;
	
	\item An empirical comparison between the capabilities of the intensity index and the antipattern metrics defined by Taba \etal \cite{Taba:2013} in the context of change prediction;
	
	\item A novel smell-aware combined change prediction model, which has more effective performance than the state of the art;
	
	\item A replication package that includes all the raw data and working data sets of our study \cite{appendix}.
	
\end{enumerate}

\noindent \textbf{Structure of the paper.} Section \ref{sec:related} discusses the related literature on change prediction models and code smell. Section \ref{sec:study} describes the design of the case study aimed at evaluating the performance of the models, while Section \ref{sec:result} reports the results achieved. Section \ref{sec:threats} discusses the threats to the validity of our empirical study. Finally, Section \ref{sec:conclusion} concludes the paper and outlines directions for future work.

\section{Related Work}
\label{sec:related}
Change-prone classes represent source code components that, for different reasons, tend to change more often than others. This phenomenon has been widely investigated by the research community \cite{khomh2012exploratory,kim:tse14,dipenta:icsm2008,bieman:metrics,soetens:2016,palomba2017diffuseness} with the aim of studying the factors contributing to the change-proneness of classes. Among all these studies, Khomh \etal \cite{khomh2012exploratory} showed that the presence of sub-optimal implementations in Java classes, \ie code smells, has a strong impact on the likelihood that such classes will be often modified by developers. The results were later confirmed by several studies in the field \cite{dambros:qsic,Olbrich10areall,palomba2017diffuseness,Spadini:icsme18}, further highlighting the relevance of code smells for change-proneness. Our work is clearly based on these findings, and aims at providing additional evidence of how code smells can be adopted in the context of prediction models having the goal of identifying change-prone classes.

Keeping track of classes that are more prone to change can be relevant for two main reasons: on the one hand, to create awareness among developers about the fact that these classes tend to change frequently; on the other hand, to highlight the presence of classes that might require preventive maintenance actions (\eg code review \cite{bacchelli:2013} or refactoring \cite{fowler:1999}) aimed at improving the quality of the source code. In this regard, previous researchers heavily investigated the feasibility of machine learning techniques for the identification of change-prone classes \cite{catolino2018enhancing,elish2013suite,eski2011empirical,girba2004yesterday,kumar2017transfer,lu2012ability,malhotra2015predicting,romano2011using,zhou2009examining}. In the following, we discuss the advances achieved in the context of change prediction models. At the same time, as this paper reports on the role of \emph{code smell intensity}, we also summarize the literature related to the detection and prioritization of code smells. 

\subsection{Change Prediction Approaches}
The most relevant body of knowledge related to change prediction techniques is represented by the use of product and process metrics as independent variables able to characterize the change-proneness of software artifacts \cite{abdi2006analyzing,arisholm2004dynamic,briand1999using,lu2012ability, malhotra2015predicting,malhotra2013investigation,tsantalis2005predicting,zhou2009examining}. Specifically, Romano \etal \cite{romano2011using} relied on code metrics for predicting change-prone the so-called fat interfaces (\ie poorly-cohesive Java interfaces), while Eski \etal \cite{eski2011empirical} proposed a model based on both CK and QMOOD metrics \cite{bansiya:2002} to estimate change-prone classes and to determine parts of the source code that should be tested first and more deeply.

Conversely, Elish \etal \cite{elish2013suite} reported the potential usefulness of process metrics for change prediction. In particular, they defined a set of \emph{evolution} metrics that describe the historical characteristics of software classes: for instance, they defined metrics like the birth date of a class or the total amount of changes applied in the past. 
As a result, their findings showed that a prediction model based on those evolution metrics can overcome the performance of structural-based techniques. These results were partially confirmed by Girba \etal \cite{girba2004yesterday}, who defined a tool that suggests change-prone code elements by summarizing previous changes. In a small-scale empirical study involving two systems, they observed that previous changes can effectively predict future modifications.

More recently, Catolino \etal \cite{catolino2018enhancing} empirically assessed the role of developer-related factors in change prediction. To this aim, they studied the performance of three developer-based prediction models relying on (i) entropy of development process \cite{hassan2009predicting}, (ii) number of developers working on a certain class \cite{bell2013limited}, and (iii) structural and semantic scattering of changes \cite{di2015role}, showing that they can be more accurate than models based on product or process metrics. Furthermore, they also defined a combined model which considers a mixture of metrics and that has shown to be up to 22\% more accurate than the previously defined ones.

Our work builds upon the findings reported above. In particular, we study to what extent the addition of information related to the presence and severity of code smells can contribute to the performance of change prediction models based on product, process, and developer-based metrics. 

Another consistent part of the state of the art concerns with the use of alternative methodologies to predict change-prone classes. For instance, the combination of (i) dependencies mined from UML diagrams and code metrics \cite{han2008behavioral,han2010measuring,rumbaugh:2004,sharafat2007probabilistic,sharafat2008change}, and (ii) genetic and learning algorithms \cite{malhotra2014new,marinescu2014good,peer2013application} have been proposed.
Finally, some studies focused on the adoption of ensemble techniques for change prediction \cite{catolino2018m,kennedy2011particle,kumar2017empirical,malhotra2017software}. In particular, Malhotra and Khanna \cite{malhotra2017software} proposed a search-based solution to the problem, adopting a Particle Swarm Optimization (PSO)-based classifier \cite{kennedy2011particle} for predicting the change-proneness of classes. The study was conducted on five Android application packages and the results encouraged the use of the adopted solution for developing change prediction models. Kumar \etal \cite{kumar2017empirical} studied the correlation between 62 software metrics and the likelihood of a class to change in the future. Afterwards, they built a change prediction model considering eight different machine learning algorithms and two ensemble techniques. The results showed that with the application of feature selection techniques, the change prediction models relying on ensemble classifiers can obtain better results. These results were partially contradicted by Catolino and Ferrucci \cite{catolino2018m}, who empirically compared the performance of three ensemble techniques (\ie, Boosting, Random Forest, and Bagging) with the one of standard machine learning classifiers (\eg, Logistic Regression) on eight open source systems. The key results of the study showed how ensemble techniques in some cases perform better than standard machine learning approaches, however the differences among them is generally small.

\subsection{Code Smell Detection and Prioritization}
Fowler defined ``bad code smells'' (shortly, ``code smells'' or simply ``smells'') as \emph{``symptoms of the presence of poor design or implementation choices applied during the development of a software system''} \cite{fowler:1999}. Starting from there, several researchers heavily investigated (i) how code smells evolve over time \cite{palomba2017scent,palomba2018large,peters2012evaluating,tufano2015and,tufano2016empirical,Tufano:icse2015}, (ii) the way developers perceive them \cite{palomba2014they,taibi2017developers,YamashitaM12}, and (iii) what is their impact on non-functional attributes of source code \cite{abbes2011empirical,Gatrell:jss,khomh2009exploratory,khomh2012exploratory,palomba2017does,palomba2017diffuseness,sjoberg2013quantifying,Yamashita:ICSE2013}. All these studies came up with a shared conclusion: code smells negatively impact program comprehension, maintainability of source code, and development costs. In the scope of this paper, the most relevant empirical studies are those reported by Khomh \etal \cite{khomh2012exploratory} and Palomba \etal \cite{palomba2017diffuseness}, who explicitly investigated the impact of code smells on software change proneness. Both the studies reported that classes affected by design flaws tend to change more frequently than classes not affected by any code smell. Moreover, refactoring practices notably help in keeping under control the change-proneness of classes. These studies are clearly those that motivate our work: indeed, following the findings on the influence of code smells, we believe that the addition of information coming from the analysis of the severity of code smells can positively improves the performance of change prediction models. As explained later in Section \ref{sec:threats}, we measure the intensity rather than the simple presence/absence of smells because a severity metric can provide us with a more fine-grained information on how much a design problem is ``dangerous'' for a certain source code class.

Starting from the findings on the negative impact of code smells on source code maintainability, the research community heavily focused on devising techniques able to automatically detect code smells. Most of these approaches rely on a two-step approach \cite{Khomh:qsic2009,Lanza:2006,Marinescu04,Moha:tse2010,Munro05-BadSmellIdentification, Oliveto10-CSMR-ABS, Tsantalis:tse2009}: in the first one, a set of structural code metrics are computed and compared against predefined thresholds; in the second one, these metrics are combined using and/or operators in order to define detection rules. If the logical relationships expressed in such detection rules are violated, a code smell is identified.
While these approaches already have good performance, Arcelli Fontana \etal \cite{Fontana:saner16} and Aniche \etal \cite{Aniche:scam16} proposed methods to further improve it by discarding false positive code smell instances or tailoring the thresholds of code metrics, respectively.

Besides structural analysis, the use of alternative sources of information for smell detection has been proposed. Ratiu \etal \cite{RatiuDGM04} and Palomba \etal \cite{palomba2013detecting, Palomba:tse2015} showed how historical information can be exploited for detecting code smells. These approaches are particularly useful when dealing with design issues arising because of evolution problems (\eg how a hierarchy evolves over time). On the other hand, Palomba \etal \cite{Palomba:icpc2016,Palomba:icsme18} adopted Information Retrieval (IR) methods \cite{Baeza-Yates:1999} to identify code smells characterized by promiscuous responsibilities (\emph{Blob} classes). 

Furthermore, Arcelli Fontana \etal \cite{ArcelliFontana:emse2016,Fontana:icsm} and Kessentini \etal \cite{boussaa:2013,Kessentini:ase2010,Kessentini-sbse:tse,Kessentini-bilevel:tosem} used machine learning and search-based algorithms to discover code smells, pointing out that a training set composed of one hundred instances is sufficient to reach very high values of accuracy. Nevertheless, Di Nucci \etal \cite{di2018detecting} recently showed that the performance of such techniques may vary depending on the exploited dataset. 
	
Finally, Morales \etal \cite{Morales:jss16} proposed a developer-based approach that leverages contextual information on the task a developer is currently working on to recommend what are the smells that can be removed on the portion of source code referring to the performed task.

In parallel with the definition of code smell detectors, several researchers faced the problem of prioritizing code smell instances based on their harmfulness for the overall maintainability of a software project. 
Vidal \etal \cite{vidal2016approach} developed a semi-automated approach that recommends a ranking of code smells based on (i) past component modifications (\eg number of changes during the system history), (ii) important modifiability scenarios, and (iii) relevance of the kind of smell assigned by developers. In a follow-up work, the same authors introduced a new criteria for prioritizing groups of code anomalies as indicators of architectural problems in evolving systems \cite{vidal2016criteria}. 

Lanza and Marinescu \cite{Lanza:2006} proposed a metric-based rules approach in order to detect code smells, or identify code problems called disharmonies. The classes (or methods) that contain a high number of disharmonies are considered more critical. Marinescu \cite{marinescu2012assessing} also presented the \emph{Flaw Impact Score}, \ie a measure of criticality of code smells that considers (i) negative influence of a code smell on coupling, cohesion, complexity, and encapsulation; (ii) granularity, namely the type of component (method or a class) that a smell affects; and (iii) severity, measured by one or more metrics analyzing the critical symptoms of the smell.

Murphy-Hill and Black \cite{murphy2010interactive} introduced an interactive visualization environment aimed at helping developers when assessing the harmfulness of code smell instances. The idea behind the tool is to visualize classes like petals, and an increase of code smell severity corresponded with an increased petal size.
Other studies exploited developers' knowledge in order to assign a level of severity with the aim to suggest relevant refactoring solutions \cite{mkaouer2014robust}, while Zhao and Hayes \cite{zhao2011rank} proposed a hierarchical approach to identify and prioritize refactoring operations based on predicted improvement to the maintainability of the software.

Besides the prioritization approaches mentioned above, more recently Arcelli Fontana and Zanoni \cite{fontana2017code} proposed the use of machine learning techniques to predict code smell severity, reporting promising results. The same authors also proposed \textsc{JCodeOdor} \cite{fontana2015towards}, a code smell detector that is able to assign a level of severity by computing the so-called \emph{intensity index}, \ie the extent to which a set of structural metrics computed on smelly classes exceed the predefined thresholds: the higher the distance between the actual and the threshold values the higher the severity of a code smell instance. As explained later in the paper (Section \ref{sec:study}), in the context of our study we adopt \textsc{JCodeOdor} since it has been previously evaluated on the dataset we exploited, reporting a high accuracy. This was therefore the best option we had to conduct our work.

\section{Research Methodology}
\label{sec:study}

In this section, we report the empirical study definition and design that we follow to test the contribution given by the addition of the code smell intensity index to existing change prediction models.

\subsection{Research Questions}
The \emph{goal} of the empirical study was to evaluate the contribution of the \emph{intensity} index in prediction models aimed at discovering change-prone classes, with the \emph{purpose} of improving the allocation of resources in preventive maintenance task such as code inspection \cite{bacchelli:2013} or refactoring \cite{fowler:1999}. The \emph{quality focus} is on the prediction performance of models that include code smell-related information when compared to state of the art approaches, while the \emph{perspective} is of researchers, who want to evaluate the effectiveness of using information about code smells when identifying change-prone components. More specifically, the empirical investigation aimed at answering the following research questions:\smallskip

\begin{itemize}
	
	\item \textbf{RQ$_1$.} \emph{To what extent does the intensity index improve the performance of existing change prediction models?}\smallskip
	
	\item \textbf{RQ$_2$.} \emph{How does the model including the intensity index as predictor compare to a model built using antipattern metrics?}\smallskip
	
	\item \textbf{RQ$_3$.} \emph{What is the gain provided by the intensity index to change prediction models when compared to other predictors?}\smallskip
	
	\item \textbf{RQ$_4$.} \emph{What is the performance of a combined change prediction model that includes smell-related information?}
	
\end{itemize}

As detailed in the next sections, the first research question ({\bf RQ$_1$}) aimed at investigating the contribution given by the intensity index within change prediction models built using different types of predictors, \ie product, process, and developer-related metrics. In {\bf RQ$_2$} we empirically compared models relying on two different types of smell-related information, \ie the intensity index \cite{fontana2015towards} and the \emph{antipattern metrics} proposed by Taba \etal \cite{Taba:2013}.
{\bf RQ$_3$} was concerned with a fine-grained analysis aimed at measuring the actual gain provided by the addition of the intensity metric within different change prediction models. Finally, {\bf RQ$_4$} had the goal to assess the performance of a change prediction model built using a combination between smell-related information and other product, process, and developer-related features.

\subsection{Context Selection}
The \emph{context} of the study was represented by a publicly available dataset coming from the \textsc{Promise} repository \cite{menzies2012promise}, which consisted of 43 releases of 14 open source software systems. Table \ref{tab:systems} reports (i) the name of the considered projects, (ii) the number of releases for each of them, (iii) their size (min-max) in terms of minimum and maximum number of classes and KLOCs across the considered releases, (iv) the percentage (min-max) of change-prone classes (identified as explained later), and (iv) the percentage (min-max) of classes affected by design problems (detected as explained later). Their selection was driven by our willingness to analyze projects having different application domains and size in order to decrease threats to external validity \cite{Ghotra:icse15,Song:tse11}. It is important to remark that the use of the \textsc{Promise} dataset required some additional effort in cleaning the data it contains: indeed, as reported by Shepperd \etal \cite{shepperd2013data}, such dataset may contain noise and/or erroneous entries that possibly negatively influence the results. To account for this aspect, before running our experiments we performed a data cleaning on the basis of the algorithm proposed by Shepperd \etal \cite{shepperd2013data}, which consists of 13 corrections able to remove identical features, features with conflicting or missing values, etc. During this step, we removed 58 entries from the original dataset. 

As for the code smells, our investigation copes with six types of design problems, namely:

\begin{itemize}
	\item \emph{God Class} (a.k.a., \emph{Blob}): A poorly cohesive class that implements different responsibilities;
	
	\item \emph{Data Class}: A class whose only purpose is holding data;
	
	\item \emph{Brain Method}: A large method implementing more than one function, being therefore poorly cohesive;
	
	\item \emph{Shotgun Surgery}: A class where every change triggers many little changes to several other classes;
	
	\item \emph{Dispersed Coupling}: A class having too many relationships with other classes of the project;
	
	\item \emph{Message Chains}: A method containing a long chain of method calls.
\end{itemize}

The choice of focusing on these specific smells was driven by two main aspects: (i) on the one hand, we took into account code smells characterizing different design problems (\eg excessive coupling vs poorly cohesive classes/methods) and having different granularities; (ii) on the other hand, as explained in the next section, we could rely on a reliable tool to properly identify and compute their intensity in the classes of the exploited dataset. 

\begin{table}[tb]
	\centering
	\caption{Characteristics of the Software Projects in Our Dataset}
	\label{tab:systems}
	%	\rowcolors{2}{gray!25}{white}
	\resizebox{\linewidth}{!}{
		\begin{tabular}{lccccc}\hline
			\textbf{System} & \textbf{Releases} & \textbf{Classes} & \textbf{KLOCs} & \textbf{\% Change Cl.} & \textbf{\% Smelly Cl.}\\\hline
			\cellcolor[gray]{0.85}Apache Ant & \cellcolor[gray]{0.85}5 & \cellcolor[gray]{0.85}83-813 & \cellcolor[gray]{0.85}20-204 & \cellcolor[gray]{0.85}24 &\cellcolor[gray]{0.85} 11-16\\
			Apache Camel & 4 & 315-571 & 70-108 & 25 & 9-14\\
			\cellcolor[gray]{0.85}Apache Forrest & \cellcolor[gray]{0.85}3 & \cellcolor[gray]{0.85}112-628 &\cellcolor[gray]{0.85}18-193 &\cellcolor[gray]{0.85}64 & \cellcolor[gray]{0.85}11-13\\
			Apache Ivy & 1 & 349 & 58 & 65 &12\\
			\cellcolor[gray]{0.85}Apache Log4j & \cellcolor[gray]{0.85}3 & \cellcolor[gray]{0.85}205-281 &\cellcolor[gray]{0.85} 38-51 &\cellcolor[gray]{0.85}26 & \cellcolor[gray]{0.85}15-19\\
			Apache Lucene & 3 & 338-2,246 & 103-466 &26 & 10-22\\
			\cellcolor[gray]{0.85}Apache Pbeans &\cellcolor[gray]{0.85}2 & \cellcolor[gray]{0.85}121-509 & \cellcolor[gray]{0.85}13-55 &\cellcolor[gray]{0.85}37 & \cellcolor[gray]{0.85}21-25\\
			Apache POI & 4 & 129-278 & 68-124 &22 & 15-19\\
			\cellcolor[gray]{0.85}Apache Synapse & \cellcolor[gray]{0.85}3 & \cellcolor[gray]{0.85}249-317 & \cellcolor[gray]{0.85}117-136 &\cellcolor[gray]{0.85}26&\cellcolor[gray]{0.85}13-17\\
			Apache Tomcat & 1 & 858 & 301 & 76 & 4\\
			\cellcolor[gray]{0.85}Apache Velocity &\cellcolor[gray]{0.85}3 & \cellcolor[gray]{0.85}229-341 & \cellcolor[gray]{0.85}57-73 &\cellcolor[gray]{0.85}26&\cellcolor[gray]{0.85} 7-13\\
			Apache Xalan & 4 & 909 & 428 & 25 & 12-22\\
			\cellcolor[gray]{0.85}Apache Xerces &\cellcolor[gray]{0.85}3 &\cellcolor[gray]{0.85} 162-736 & \cellcolor[gray]{0.85}62-201 & \cellcolor[gray]{0.85}24 & \cellcolor[gray]{0.85}5-9\\
			JEdit & 5 & 228-520 & 39-166 & 23 & 14-22\\\hline
		\end{tabular}
	}
\end{table}

\subsection{\textbf{RQ$_1$} - The contribution of the Intensity Index}\label{sec:model}
To answer our first research question, we needed to (i) identify code smells in the subject projects and compute their intensity, and (ii) select a set of existing change prediction models where to add the information on the intensity of code smells. Furthermore, we proceeded with the training and testing of the built change prediction models. The following subsections detail the process conducted to perform such steps.\smallskip

\begin{table*}[!htbp]
	\centering
	\caption{Code Smell Detection Strategies (the complete names of the metrics are given in Table~\ref{tab:metrics} and the explanation of the rules in Table~\ref{tab:csrules_rationale})}
	\label{tab:csrules}
	\resizebox{1\linewidth}{!}{
		\renewcommand{\arraystretch}{1.5}
		\footnotesize
		\setlength{\tabcolsep}{0.3em}
		\begin{tabularx}{\linewidth}{lX}
			\toprule
			Code Smells	& Detection Strategies: LABEL($n$) $\rightarrow$ LABEL has value $n$ for that smell\\
			\midrule
			God Class	& LOCNAMM $\geq$ HIGH(176) $\wedge$ WMCNAMM $\geq$ MEAN(22) $\wedge$ NOMNAMM $\geq$ HIGH(18) $\wedge$ TCC $\leq$ LOW(0.33) $\wedge$ ATFD $\geq$ MEAN(6)\\
			%\midrule
			Data Class	& WMCNAMM $\leq$ LOW(14) $\wedge$ WOC $\leq$ LOW(0.33) $\wedge$	NOAM $\geq$ MEAN(4) $\wedge$ NOPA $\geq$ MEAN(3)\\
			%\midrule
			Brain Method	& (LOC $\geq$ HIGH(33) $\wedge$ CYCLO $\geq$ HIGH(7) $\wedge$ MAXNESTING $\geq$ HIGH(6)) $\vee$	(NOLV $\geq$ MEAN(6) $\wedge$ ATLD $\geq$ MEAN(5))\\
			%\midrule
			Shotgun Surgery & CC $\geq$ HIGH(5) $\wedge$ CM $\geq$ HIGH(6) $\wedge$ FANOUT $\geq$ LOW(3)\\
			%\midrule
			Dispersed Coupling	& CINT $\geq$ HIGH(8) $\wedge$ CDISP $\geq$ HIGH(0.66)\\
			%\midrule
			Message Chains	& MaMCL $\geq$ MEAN(3) $\vee$ (NMCS $\geq$ MEAN(3) $\wedge$ MeMCL $\geq$ LOW(2))\\
			\bottomrule
			%\multicolumn{3}{l}{\emph{Legend}: LABEL($n$) means that LABEL assumes value $n$ for that smell}\\
		\end{tabularx}
		\renewcommand{\arraystretch}{1}
	}
\end{table*}

\begin{table*}[!htbp]
	\centering
	\caption{Metrics used for Code Smells Detection}\label{tab:metrics}
	\resizebox{1\linewidth}{!}{
		\footnotesize
		\setlength{\tabcolsep}{0.3em}
		\begin{tabularx}{\linewidth}{lp{25mm}p{70mm}}
			\toprule
			Short Name	    & Long Name & Definition\\
			\midrule
			ATFD		    & Access To Foreign Data    & The number of attributes from unrelated classes belonging to the system, accessed directly or by invoking accessor methods.\\\hline
			ATLD	& Access To Local Data      & The number of attributes declared by the current classes accessed by the measured method directly or by invoking accessor methods.\\\hline
			CC			    & Changing Classes          & The number of classes in which the methods that call the measured method are defined in.\\\hline
			CDISP 		    & Coupling Dispersion       & The number of classes in which the operations called from the measured operation are defined, divided by CINT.\\\hline
			CINT		    & Coupling Intensity        & The number of distinct operations called by the measured operation.\\\hline
			CM			    & Changing Methods          & The number of distinct methods that call the measured method.\\\hline
			CYCLO 		    & McCabe Cyclomatic Complexity  & The maximum number of linearly independent paths in a method. A path is linear if there is no branch in the execution flow of the corresponding code.\\\hline
			FANOUT		    &                               & Number of called classes.\\\hline
			LOC 		    & Lines Of Code             & The number of lines of code of an operation or of a class, including blank lines and comments.\\\hline
			LOCNAMM & Lines of Code Without Accessor or Mutator Methods & The number of lines of code of a class, including blank lines and comments and excluding accessor and mutator methods and corresponding comments.\\\hline
			MaMCL	& Maximum Message Chain Length  & The maximum length of chained calls in a method.\\\hline
			MAXNESTING	    & Maximum Nesting Level         & The maximum nesting level of control structures within an operation.\\\hline
			MeMCL	& Mean Message Chain Length     & The average length of chained calls in a method.\\\hline
			NMCS 	& Number of Message Chain Statements    & The number of different chained calls in a method.\\\hline
			NOAM		    & Number Of Accessor Methods    & The number of accessor (getter and setter) methods of a class.\\\hline
			NOLV 		    & Number Of Local Variables     & Number of local variables declared in a method. The method's parameters are considered local variables.\\\hline
			NOMNAMM	& Number of Not Accessor or Mutator Methods & The number of methods defined locally in a class, counting public as well as private methods, exclud-ing accessor or mutator methods.\\\hline
			\emph{*}NOMNAMM	& Number of Not Accessor or Mutator Methods & The number of methods defined locally in a class, counting public as well as private methods, excluding accessor or mutator methods.\\\hline
			NOPA 		    & Number Of Public Attributes   & The number of public attributes of a class.\\\hline
			TCC			    & Tight Class Cohesion          & The normalized ratio between the number of methods directly connected with other methods through an instance variable and the total number of possible connections between methods. A direct connection between two methods exists if both access the same instance variable directly or indirectly through a method call. TCC takes its value in the range [0,1].\\\hline
			WMCNAMM	& Weighted Methods Count of Not Accessor or Mutator Methods & The sum of complexity of the methods that are defined in the class, and are not accessor or mutator methods. We compute the complexity with the Cyclomatic Complexity metric (CYCLO).\\\hline
			WOC 		    & Weight Of Class               & The number of ``functional'' (\ie non-abstract, non-accessor, non-mutator) public methods divided by the total number of public members.\\\hline
			\bottomrule
		\end{tabularx}
	}
\end{table*}

\begin{table*}[!htbp]
	\centering
	\caption{Code Smell Detection Rationale and Details}
	\label{tab:csrules_rationale}
	\resizebox{1\linewidth}{!}{
		\footnotesize
		\setlength{\tabcolsep}{0.3em}
		\begin{tabularx}{\linewidth}{llX}
			\toprule
			& Clause  & Rationale\\
			\midrule
			\multirow{10}{*}{\begin{sideways}God Class\end{sideways}}
			& LOCNAMM $\geq$ HIGH & \emph{Too much code.}
			We use LOCNAMM instead of LOC, because getter and setter methods are often generated by the IDE. A class that has getter and setter methods, and a class that has not getter and setter methods, must have the same ``probability'' to be detected as God Class.\\ 
			& WMCNAMM $\geq$ MEAN & \emph{Too much work and complex.}
			Each method has a minimum cyclomatic complexity of one, hence also getter and setter add cyclomatic complexity to the class. We decide to use a complexity metric that excludes them from the computation.\\
			& NOMNAMM $\geq$ HIGH & \emph{Implements a high number of functions.}
			We exclude getter and setter because we consider only the methods that effectively implement functionality of the class.\\
			& TCC $\leq$ LOW & \emph{Functions accomplish different tasks.}\\
			& ATFD $\geq$ MEAN & \emph{Uses many data from other classes.}\\
			\midrule
			\multirow{8}{*}{\begin{sideways}Data Class\end{sideways}}
			& WMCNAMM $\leq$ LOW & \emph{Methods are not complex.}
			Each method has a minimum cyclomatic complexity of one, hence also getter and setter add cyclomatic complexity to the class. We decide to use a complexity metric that exclude them from the computation.\\
			& WOC $\leq$ LOW & \emph{The class offers few functionalities.}
			This metrics is computed as the number of functional (non-accessor) public methods, divided by the total number of public methods. A low value for the WOC metric means that the class offers few functionalities.\\
			& NOAM $\geq$ MEAN & \emph{The class has many accessor methods.} \\ 
			& NOPA $\geq$ MEAN &  \emph{The class has many public attributes.}\\
			\midrule
			\multirow{6}{*}{\begin{sideways}Brain Method\end{sideways}}
			& LOC $\geq$ HIGH & \emph{Too much code.} \\
			& CYCLO $\geq$ HIGH & \emph{High functional complexity} \\ 
			& MAXNESTING $\geq$ HIGH & \emph{High functional complexity. Difficult to understand.}\\ 
			& NOLV $\geq$ MEAN & \emph{Difficult to understand.}
			More the number of local variable, more the method is difficult to understand.\\
			& ATLD $\geq$ MEAN & \emph{Uses many of the data of the class.}
			More the number of attributes of the class the method uses, more the method is difficult to understand.\\
			\midrule
			\multirow{5}{*}{\begin{sideways}Shot. Surg.\end{sideways}}
			& CC $\geq$ HIGH & \emph{Many classes call the method.} \\
			& CM $\geq$ HIGH & \emph{Many methods to change.}\\ 
			& FANOUT $\geq$ LOW & \emph{The method is subject to being changed.}
			If a method interacts with other classes, it is not a trivial one. We use the FANOUT metric to refer Shotgun Surgery only to those methods that are more subject to be changed. We exclude for example most of the getter and setter methods.\\
			\midrule
			\multirow{4}{*}{\begin{sideways}Dis. Coup.\end{sideways}}
			& CINT $\geq$ HIGH &  \emph{The method calls too many other methods.}
			With CINT metric, we measure the number of distinct methods called from the measured method.\\
			& CDISP $\geq$ HIGH & \emph{Calls are dispersed in many classes.}
			With CDISP metric, we measure the dispersion of called methods: the number of classes in which the methods called from the measured method are defined in, divided by CINT.\\
			\midrule
			\multirow{7}{*}{\begin{sideways}Mess. Chain\end{sideways}}
			& MaMCL $\geq$ MEAN & \emph{Maximum Message Chain Length.}
			A Message Chains has a minimum length of two chained calls, because a single call is trivial. We use the MaMCL metric to find out the methods that have at least one chained call with a length greater than the mean.\\
			& NMCS $\geq$ MEAN &  \emph{Number of Message Chain Statements.}
			There can be more Message Chain Statement: different chains of call. More the number of Message Chain Statements, more the method is interesting respect to Message Chains code smell.\\
			& MeMCL $\geq$ LOW & \emph{Mean of Message Chain Length.}
			We would find out non-trivial Message Chains, so we need always to check against the Message Chain Statement length.\\
			\bottomrule
			%\multicolumn{3}{l}{\emph{Legend}: LABEL($n$) means that LABEL assumes value $n$ for that smell}\\
		\end{tabularx}
	}
\end{table*}

\subsubsection{Code Smell Intensity Computation}
To compute the severity of code smells in the context of our work we employed \textsc{JCodeOdor} \cite{fontana2015towards}, a tool that is able to both identify code smells and assign to them a degree of severity by computing the so-called \emph{intensity index}. Such an index is represented by a real number contained in the range [1, 10]. Before reporting the detailed steps adopted by \textsc{JCodeOdor} to compute the intensity index, it is worth remarking that our choice to use this tool was driven by two observations. In the first place, \textsc{JCodeOdor} works with all the code smells considered in our work: in literature it is the prioritization approach that deal with the highest number of code smells (6 vs the 5 treated by Vidal \etal \cite{vidal2016approach}). Moreover, it is fully automated, meaning that it does not require any human intervention while computing the intensity of code smells. Finally, it is highly accurate: in a previous work by Palomba \etal \cite{palomba2017toward} the tool was empirically assessed on the same dataset adopted in our context, showing an F-Measure of 80\%. For these reasons, we believe that \textsc{JCodeOdor} was the best option we had to conduct our study.

From a technical point of view, given the set of classes composing a certain software system the tool performs two basic steps to compute the intensity of code smells:

\begin{enumerate}
	
	\item \emph{Detection Phase.} Given a software system as input, the tool starts by detecting code smells relying on the detection strategies reported in Table~\ref{tab:csrules}. Basically, each strategy is represented by a logical composition of predicates, and each predicate is based on an operator that compares a metric with a threshold~\cite{Lanza:2006, bookchapert:Palomba}. 
	Such detection strategies are similar to those defined by Lanza and Marinescu~\cite{Lanza:2006}, who used the set of code metrics described in Table~\ref{tab:metrics} to identify the six code smell types in our study. To ease the comprehension of the detection approach, Table~\ref{tab:csrules_rationale} describes the rationale behind the use of each predicate of the detection strategies.
	
	On the basis of these detection rules, a class/method of a project is marked as \emph{smelly} if one of the logical propositions shown in Table~\ref{tab:csrules} is true, \ie if the actual metrics computed on the class/method exceed the threshold values defined in the detection strategy. It is worth pointing out that the thresholds used by \textsc{JCodeOdor} were empirically calibrated on 74 systems belonging to the \textsc{Qualitas Corpus} dataset~\cite{Tempero2010} and are derived from the statistical distribution of the metrics contained in the dataset~\cite{fontana2015towards,fontana2015automatic}. 
	
	\smallskip
	\item \emph{Intensity Computation.} If a class/method is identified by the tool as smelly, the actual value of a given metric used for the detection will exceed the threshold value, and it will correspond to a percentile value on the metric distribution placed between the threshold and the maximum observed value of the metric in the system under analysis. The placement of the actual metric value in that range represents the ``exceeding amount'' of a metric with respect to the defined threshold. Such ``exceeding amounts'' are then normalized in the range [1,10] using a min-max normalization process \cite{theodoridis2008pattern}: specifically, this is a feature scaling technique where the values of a numeric range are reduced to a scale between 1 and 10. To compute $z$, \ie the normalized value, the following formula is applied:
	
	\begin{equation}
	z = [\frac{x - min(x)}{max(x) - min(x)}] \cdot 10
	\end{equation}
	
	where $min$ and $max$ are the minimum and maximum values observed in the distribution. This step allows to have the ``exceeding amount'' of each metric in the same scale. To have a unique value representing the intensity of the code smell affecting the class, the mean of the normalized ``exceeding amounts'' is computed.
	
\end{enumerate}
\medskip

\subsubsection{Selection of Basic Prediction Models}
\label{sec:basicModels}
Our conjecture was concerned with the gain given by the addition of information on the intensity of code smells within existing change prediction models. To test such a conjecture, we needed to identify the state of the art techniques where to add the intensity index: we selected three models based on product, process, and developer-related metrics that were shown to be accurate in the context of change prediction \cite{catolino2018enhancing,di2015role,elish2013suite,zhou2009examining}. \smallskip

\noindent \textbf{Product Metrics-based Model.} The first basic baseline is represented by the change prediction model devised by Zhou \etal \cite{zhou2009examining}. It is composed of a set of metrics computed on the basis of the structural properties of source code: these are cohesion (\ie the Lack of Cohesion of Method --- LCOM), coupling (\ie the Coupling Between Objects --- CBO --- and the Response for a Class --- RFC), and inheritance metrics (\ie the Depth of Inheritance Tree --- DIT). To actually compute these metrics, we relied on a publicly available tool originally developed by Spinellis \cite{spinellis2005tool}. In the following, we refer to this model as SM, \ie Structural Model. \medskip

\noindent \textbf{Process Metrics-based Model.} In their study, Elish \etal \cite{elish2013suite} reported that process metrics can be exploited as better predictors of change-proneness with respect to structural metrics. For this reason, our second baseline was the Evolution Model (EM) proposed by Elish \etal \cite{elish2013suite}. More specifically, this model relies on the metrics shown in Table \ref{tab:Metrics}, which capture different aspects of the evolution of classes, \eg the weighted frequency of changes or the first time changes introduced. To compute these metrics, we adopted the tool that was previously developed by Catolino \etal \cite{catolino2018enhancing}. In the following, we refer to this model as PM, \ie Process Model. \medskip

\begin{table}[tb]
	\centering
	\caption{Independent variables considered by Elish et al.}
	\label{tab:Metrics}
	\resizebox{0.7\linewidth}{!}{
		\begin{tabular}{p{2.2cm}p{5.5cm}}
			\hline
			Acronym&Metric\\
			\hline			
			\cellcolor[gray]{0.85}BOC&\cellcolor[gray]{0.85}Birth of a Class\\
			FCH & First Time Changes Introduced to a Class \\
			\cellcolor[gray]{0.85}FRCH&\cellcolor[gray]{0.85}Frequency of Changes\\
			LCH&Last Time Changes Introduced to a Class\\
			%\cellcolor[gray]{0.85}WCH&\cellcolor[gray]{0.85}Halstead Complexity Measures\\
			\cellcolor[gray]{0.85}WCD&\cellcolor[gray]{0.85}Weighted Change Density\\
			WFR&Weighted Frequency of Changes\\
			\cellcolor[gray]{0.85}TACH&\cellcolor[gray]{0.85}Total Amount of Changes\\
			ATAF&Aggregated Change Size Normalized by Frequency of Change\\
			\cellcolor[gray]{0.85}CHD&\cellcolor[gray]{0.85}Change Density\\
			LCA&Last Change Amount\\
			\cellcolor[gray]{0.85}LCD&\cellcolor[gray]{0.85}Last Change Density\\
			CSB&Changes since the Birth\\
			\cellcolor[gray]{0.85}CSBS&\cellcolor[gray]{0.85}Changes since the Birth Normalized by Size\\
			ACDF&Aggregated Change Density Normalized by Frequency of Change\\
			\cellcolor[gray]{0.85}CHO&\cellcolor[gray]{0.85}Change Occurred\\
			\hline
		\end{tabular}
	}
\end{table}

\noindent \textbf{Developer-Related Model.} In our previous work \cite{catolino2018enhancing}, we demonstrated how developer-related factors can be exploited within change prediction models since they provide orthogonal information with respect to product and process metrics that takes into account how developers perform modifications and how complex the development process is. Among all the available developer-based models developed in literature~\cite{bell2013limited,di2015role,hassan2009predicting}, in this paper we relied on the Developer Changes Based Model (DCBM) devised by Di Nucci \etal \cite{di2015role}, as it was shown to be the most effective one in the context of change prediction. Such a model uses as predictors the so-called structural and semantic scattering of the developers that worked on a code component in a given time period $\alpha$. Specifically, for each class $c$, the two metrics are computed as follows:

\begin{gather}
\mbox{StrScatPred}_{c,\alpha} = \sum_{d \in developers_{c,\alpha}} \mbox{StrScat}_{d,\alpha}\\
\mbox{SemScatPred}_{c,\alpha} = \sum_{d \in developers_{c,\alpha}} \mbox{SemScat}_{d,\alpha}
\end{gather}

\noindent where $developers_{c,\alpha}$ represents the set of developers that worked on the class $c$ during a certain period  $\alpha$, and the functions $StrScat_{d,\alpha}$ and $SemScat_{d,\alpha}$ return the structural and semantic scattering, respectively, of a developer $d$ in the time window $\alpha$. 
Given the set $CH_{d,\alpha}$ of classes changed by a developer $d$ during a certain period $\alpha$, the formula of structural scattering of a developer is:

\begin{equation}
\mbox{StrScat}_{d,\alpha} = |CH_{d,\alpha}| \times \underset{\forall c_i,c_j \in CH_{d,\alpha}}{average} [dist(c_i,c_j)]
\end{equation}

\noindent where $dist$ is the distance in number of packages from class $c_{i}$ to class $c_{j}$. The structural scattering is computed by applying the shortest path algorithm on the graph representing the system’s package structure.
Regarding the semantic scattering of a developer, it is based on the textual similarity of the classes changed by a developer in a certain period $\alpha$ and it is computed as:
\begin{equation}
\mbox{SemScat}_{d,\alpha} = |CH_{d,\alpha}| \times \frac{1}{\underset{\forall c_i,c_j \in CH_{d,\alpha}}{average} [sim(c_i,c_j)]}
\end{equation}
\noindent where the $sim$ function returns the textual similarity between the classes $c_i$ and $c_j$ according to the measurement performed using the \emph{Vector Space Model} (VSM) \cite{baeza1999modern}. The metric ranges between zero (no textual similarity) and one (the textual content of the two classes is identical). In our study, we set the parameter $\alpha$ of the approach as the time window between two releases $R-1$ and $R$, as done in previous work \cite{palomba2017toward}.\medskip

\noindent It is important to note that all the baseline models might be affected by multi-collinearity \cite{Obrien2007}, which occurs when two or more independent variables are highly correlated and can be predicted one from the other, thus possibly leading to a decrease of the prediction capabilities of the resulting model \cite{Shepperd:tseResearcherBias,Tantithamthavorn:comments}. For this reason, we decided to use the \emph{vif} (variance inflation factors) function \cite{Obrien2007} implemented in R\footnote{http://cran.r-project.org/web/packages/car/index.html} to discard non-relevant variables. 
\emph{Vif} is based on the square of the multiple correlation coefficient resulting from regressing a predictor variable against all other predictor variables. If a variable has a strong linear relationship with at least one other variable, the correlation coefficient would be close to 1, and VIF for that variable would be large. A VIF greater than 10 is a signal that the model has a collinearity problem.

	\begin{table}[!htbp]
	\centering
	\caption{Changes extracted by \textsc{ChangeDistiller} while computing the change-proneness. `\checkmark' symbols indicate the types we considered in our study.}
	\label{tab:changeDistiller}
	\resizebox{0.5\linewidth}{!}{
		\begin{tabular}{lc}
			\rowcolor{gray20}
			ChangeDistiller & Our Study\\
			
			\cellcolor[gray]{0}\textcolor{white}{{\textbf{Statement-level changes}}} & \cellcolor[gray]{0}\\
			
			\rowcolor[HTML]{EFEFEF} 
			Statement Ordering Change & \checkmark \\
			Statement Parent Change & \checkmark \\
			\rowcolor[HTML]{EFEFEF}
			Statement Insert & \checkmark \\
			Statement Delete & \checkmark \\
			\rowcolor[HTML]{EFEFEF}
			Statement Update & \checkmark \\
			
			\cellcolor[gray]{0}\textcolor{white}{{\textbf{Class-body changes}}} & \cellcolor[gray]{0}\\
			
			\rowcolor[HTML]{EFEFEF} 
			Insert attribute & \checkmark \\
			Delete attribute & \checkmark \\
			
			\cellcolor[gray]{0}\textcolor{white}{{\textbf{Declaration-part changes}}} & \cellcolor[gray]{0}\\
			
			\rowcolor[HTML]{EFEFEF} 
			Access modifier update & \checkmark \\
			Final modifier update & \checkmark \\
			
			\cellcolor[gray]{0}\textcolor{white}{{\textbf{Declaration-part changes}}} & \cellcolor[gray]{0}\\
			
			\rowcolor[HTML]{EFEFEF} 
			Increasing accessibility change & \checkmark \\
			Decreasing accessibility change & \checkmark \\
			\rowcolor[HTML]{EFEFEF} 
			Final Modified Insert & \checkmark \\
			Final Modified Delete & \checkmark \\
			
			\cellcolor[gray]{0}\textcolor{white}{{\textbf{Attribute declaration changes}}} & \cellcolor[gray]{0}\\
			
			\rowcolor[HTML]{EFEFEF} 
			Attribute type change & \checkmark \\
			Attribute renaming change & \checkmark \\
			
			\cellcolor[gray]{0}\textcolor{white}{{\textbf{Method declaration changes}}} & \cellcolor[gray]{0}\\
			
			\rowcolor[HTML]{EFEFEF} 
			Return type insert & \checkmark \\
			Return type delete & \checkmark \\
			\rowcolor[HTML]{EFEFEF} 
			Return type update & \checkmark \\
			Method renaming & \checkmark \\
			\rowcolor[HTML]{EFEFEF} 
			Parameter insert & \checkmark \\
			Parameter delete & \checkmark \\
			\rowcolor[HTML]{EFEFEF} 
			Parameter ordering change & \checkmark \\
			Parameter renaming & \checkmark \\
			
			\cellcolor[gray]{0}\textcolor{white}{{\textbf{Class declaration changes}}} & \cellcolor[gray]{0}\\
			
			\rowcolor[HTML]{EFEFEF} 
			Class renaming & \checkmark \\
			Parent class insert & \checkmark \\
			\rowcolor[HTML]{EFEFEF} 
			Parent class delete & \checkmark \\
			Parent class update & \checkmark \\\hline		
		\end{tabular}
	}
\end{table}

\subsubsection{Dependent Variable} 
Our dependent variable is represented by the actual change-proneness of the classes in our dataset. To compute it, we followed the guidelines provided by Romano \etal \cite{romano2011using}, who considered a class as change-prone if, in a given time period \emph{TW}, it underwent a number of changes higher than the median of the distribution of the number of changes experienced by all the classes of the system. In particular, for each pair of commits ($c_i$ , $c{i+1}$) of \emph{TW} we run \textsc{ChangeDistiller} \cite{fluri2007change}, a tree differencing algorithm able to extract the fine-grained code changes between $c_i$ and $c{i+1}$. Table \ref{tab:changeDistiller} reports the entire list of change types identified by the tool. As it is possible to observe, we considered all of them while computing the number of changes. It is worth mentioning that the tool ignores white space-related differences and documentation-related updates: in this way, it only considers the changes actually applied on the source code. More importantly, \textsc{ChangeDistiller} is able to identify rename refactoring operations: this means that we could handle cases where a class was modified during the change history, thus not biasing the correct counting of the number of changes. The dataset with the oracle is available on the online appendix \cite{appendix}.\smallskip

\subsubsection{Experimented Machine Learning Models}
Once we had defined dependent and independent variables of interest, we could finally build machine learning models. As we were interested in understanding and measuring the contribution given by the intensity within the three baselines, we built two types of prediction models for each baseline: the first type of model \emph{does not include} the intensity index as predictor, thus relying on the original features only; the second type of model that \emph{includes} the intensity index as an additional predictor. Using this procedure, we experimented with 6 different models, and we could control the actual amount of improvement given by the intensity index with respect to the baselines (if any). It is worth remarking that, for \emph{non-smelly} classes, the intensity value is set to 0.

\subsubsection{Classifier Selection}
The final step of our prediction model construction methodology was concerned with the selection of the best machine learning classifier able to distinguish change-prone and non-change-prone classes. The current literature proposed several alternatives (\eg Romano and Pinzger~\cite{romano2011using} adopted Support Vector Machines~\cite{bottou:1992}, while Tsantalis \etal \cite{tsantalis2005predicting} relied on Logistic Regression \cite{le1992ridge}), and thus there is not a bullet-proof solution that ensures the best overall performance. For this reason, in our work we experimented with different classifiers, \ie ADTree \cite{Freund:1999}, Decision Table Majority \cite{Kohavi:1995}, Logistic Regression \cite{leCessie:1992}, Multilayer Perceptron \cite{Rosenblatt:1961}, Naive Bayes \cite{John:1995}, and Simple Logistic Regression\cite{Chao}. To select the right classifier to use in our situation, we empirically compared the results achieved when applying each classifier on each experimented  model on the software systems in our study. Overall, the best results were obtained using the Simple Logistic Regression. In the remaining of the paper, we only report the results obtained when using this classifier, while a complete report of the performance of other classifiers is available in our online appendix \cite{appendix}.

\subsubsection{Validation Strategy}
As for validation strategy we adopted \emph{10-Fold Cross Validation} \cite{stone1974cross}. This methodology randomly partitions the data into 10 folds of equal size, applying a stratified sampling. A single fold is used as test set, while the remaining ones are used as training set. The process was repeated 10 times, using each time a different fold as test set. Then, the model performances were reported using the mean achieved over the ten runs. It is important to note that we repeated the 10-fold validation 100 times (each time with a different seed) to cope with the randomness arising from using different data splits \cite{HallBBGC11}. %We adopted this strategy since it has been widely used in the context of change prediction \cite{kumar2017empirical,catolino2018enhancing,elish2013suite}.\smallskip

\subsubsection{Evaluation Metrics}\label{subsec:evaluationMetrics}
To measure and compare the performance of the models, we started computing two well-known metrics such as \emph{precision} and \emph{recall} \cite{BaezaYates}, which are defined as follow:
\begin{equation}
\mathit{precision} = {{TP} \over {TP+FP}}
\qquad
\mathit{recall} = {{TP} \over {TP+TN}}
\end{equation}

where \emph{TP} is the number of true positives, \emph{TN} the number of true negatives, and \emph{FP} the number of false positives. 
In the second place, to have a unique value representing the goodness of the model, we computed the F-Measure, \ie the harmonic mean of precision and recall:
\begin{equation}
F\mbox{-}Measure = 2 * \frac{precision*recall}{precision+recall}
\end{equation}

Moreover, we considered another indicator: the Area Under the ROC Curve (AUC-ROC) metric. This measure quantifies the overall ability of a change prediction model to discriminate between change-prone and non-change-prone classes: the closer the AUC-ROC to 1 the higher the ability of the classifier, while the closer the AUC-ROC to 0.5 the lower its accuracy. In other words, this metric can quantity how rubust the model is when discriminating the two binary classes.

In addition, we compared the performance achieved by the experimented prediction models from a statistical point of view. As we performed comparisons over multiple datasets, we employed the Scott-Knott Effect Size Difference (ESD) test \cite{Tantithamthavorn:icse2016,Tantithamthavorn:2017}, considering the AUC-ROC that the different models obtained over the considered systems. This test represents an effect-size aware variant of Scott-Knott test \cite{ScottKnott}: differently from the original one, it (i) hierarchically clusters the set of treatment means into statistically distinct groups, (ii) corrects the non-normal distribution of a dataset if needed, and (iii) merges two statistically distinct groups in case their effect size---measured using Cliff's Delta (or $d$) \cite{Cliff:2005}---is negligible, so that the creation of trivial groups is avoided. To perform the test, we relied on the implementation provided by Tantithamthavorn \etal \cite{Tantithamthavorn:2017}.
%The rationale behind the usage of this test was that the Scott-Knott ESD can be adopted to control dataset-specific performances: indeed, it evaluates the performances of the different prediction models on each dataset in isolation, thus ranking the top models based on their performances on each project. For this reason, we had 43 different Scott-Knott ranks that we analyzed by measuring the likelihood of a model to be in the top Scott-Knott ESD rank, as done in previous work \cite{Tantithamthavorn:2017,Ghotra:2017}\smallskip

\subsection{\textbf{RQ$_2$} - Comparison between Intensity Index and Antipattern Metrics}
In {\bf RQ$_2$} our goal was to compare the performance of change prediction models relying on the intensity index against the one achieved by models exploiting other existing smell-related metrics. In particular, the comparison was done considering the so-called \emph{antipattern} metrics, which were defined by Taba \etal \cite{Taba:2013}: these are three metrics aimed at capturing different aspects related to the maintainability of classes affected by code smells. More specifically:

\begin{itemize}
	
	\item the Average Number of Antipatterns (ANA) computes how many code smells there were in the previous releases of a class over the total number of releases. This metric is based on the assumption that classes that have been more prone to be smelly in the past are somehow more prone to be smelly in the future;
	
	\smallskip
	\item the Antipattern Complexity Metric (ACM) computes the entropy of changes involving smelly classes. Such entropy refers to the one originally defined by Hassan \cite{hassan2009predicting} in the context of defect prediction. The conjecture behind its use relates to the fact that a more complex development process might lead to the introduction of code smells;
	
	\smallskip
	\item the Antipattern Recurrence Length (ARL) measures the total number of subsequent releases in which a class has been affected by a smell. This metric relies on the same underlying conjecture as ANA, \ie the more a class has been smelly in the past the more it will be smelly in the future.
	
\end{itemize}

To compute these metrics, we employed the tool developed by Palomba \etal \cite{palomba2017toward}. Then, as done in the context of \textbf{RQ$_1$}, we plugged the \emph{antipattern} metrics into the experimented baselines and assess the performance of the resulting change prediction models using the same set of evaluation metrics described in Section \ref{subsec:evaluationMetrics}, \ie F-Measure and AUC-ROC. Finally, we statistically compared such performance with the one obtained by the models including the intensity index as predictor. 

Besides the comparison in terms of evaluation metrics, we also analyzed the extent to which the two types of models are complementary with respect to the classification of change-prone classes. This was done with the aim of assessing whether the two models, relying on different smell-related information, can correctly identify the change-proneness of different classes. More formally, let $m_{int}$ be the model built plugging in the intensity index; let $m_{ant}$ be the model built by considering the antipattern metrics, we computed the following overlap metrics on the set of \emph{smelly and change-prone} instances of each system:\smallskip

\begin{equation}
TP_{m_{int} \cap m_{ant}} = {|TP_{m_{int}} \cap TP_{m_{ant}}| \over |TP_{m_{int}} \cup TP_{m_{ant}}|} \% 
\end{equation}
\smallskip
\begin{equation}
TP_{m_{int} \setminus m_{ant}} = {|TP_{m_{int}} \setminus TP_{m_{ant}}| \over |TP_{m_{int}} \cup TP_{m_{ant}}|} \%
\end{equation}
\smallskip
\begin{equation}
TP_{m_{ant} \setminus m_{int}} = {|TP_{m_{ant}} \setminus TP_{m_{int}}| \over |TP_{m_{ant}} \cup TP_{m_{int}}|} \%
\end{equation}

\smallskip
\noindent where $TP_{m_{int}}$ represents the set of change-prone classes correctly classified by the prediction model $m_{int}$, while $TP_{m_{ant}}$ is the set of change-prone classes correctly classified by the prediction model $m_{ant}$. The $TP_{m_{int} \cap m_{ant}}$ metric measures the overlap between the sets of true positives correctly identified by both models $m_{int}$ and $m_{ant}$, $TP_{m_{int} \setminus m_{ant}}$ measures the percentage of change-prone classes correctly classified by $m_{int}$ only and missed by $m_{ant}$, and $TP_{m_{ant} \setminus m_{int}}$ measures the percentage of change-prone classes correctly classified by $m_{ant}$ only and missed by $m_{int}$. \smallskip

\subsection{\textbf{RQ$_3$} - Gain Provided by the Intensity Index}
In {\bf RQ$_3$} we conducted a \emph{fine-grained} investigation aimed at measuring how important is the intensity index with respect to the other features (\ie product, process, developer-related, and \emph{antipattern} metrics) composing the experimented models. To this aim, we used an \emph{information gain} algorithm~\cite{Quinlan:1986} to quantify the gain provided by adding the intensity index in each prediction model. In our context, this algorithm ranked the features of the models according to their ability to predict the change-proneness of classes. More specifically, let $M$ be a change prediction model, let $P=\left\{p_1,\dots, p_n\right\}$ be the set of predictors composing $M$, an \emph{information gain} algorithm \cite{Quinlan:1986} applies the following formula to compute a measure which defines the difference in entropy from before to after the set $M$ is split on an attribute $p_i$:

\begin{equation}
InfoGain (M, p_i) = H(M) - H(M | p_i)
\end{equation}
where the function $H(M)$ indicates the entropy of the model that includes the predictor $p_i$, while the function $H(M | p_i)$ measures the entropy of the model that does not include $p_i$. Entropy is computed as follow:

\begin{equation}
H(M) = - \sum_{i=1}^{n} prob(p_i) \log_2 prob(p_i)
\end{equation}

From a more practical perspective, the algorithm quantifies how much uncertainty in $M$ was reduced after splitting $M$ on predictor $p_i$. In our work, we employed the \emph{Gain Ratio Feature Evaluation} algorithm \cite{Quinlan:1986} implemented in the \textsc{Weka} toolkit \cite{Weka}, which ranks $p_1,\dots, p_n$ in descending order based on the contribution provided by $p_i$ to the decisions made by $M$. In particular, the output of the algorithm is a ranked list in which the predictors having the higher expected reduction in entropy are placed on the top. Using this procedure, we evaluated the relevance of the predictors in the change prediction models experimented, possibly understanding whether the addition of the intensity index gives a higher contribution with respect to the structural metrics from which it is derived (\ie metrics used for the detection of the smells) or with respect the other metrics contained in the models. 

\subsection{\textbf{RQ$_4$} - Combining All Predictors and Smell-Related Information}
As a final step of our study, we aimed to study the possibility to devise a combined model able to mix together standard change-proneness predictors (\ie structural, process, and developer-related metrics) and smell-related information to achieve better prediction performance. To do it, we firstly put all the independent variables considered in the study in a single dataset, thus putting them all together. In the second place, we applied the variable removal procedure based on the \emph{vif} function (see Section \ref{sec:basicModels} for details on this technique): in this way, we were able to remove the independent variables that do not significantly influence the performance of the combined model. Finally, we tested the ability of the newly devised model using the same procedures and metrics used in the context of {\bf RQ$_1$}, \ie F-measure, AUC-ROC, and Brier score, and statistically comparing the performance of the experimented models by means of Scott-Knott ESD test.

\section{Analysis of the Results}
\label{sec:result}
% !TEX root =  main.tex

\begin{figure*}[h]\centering
	\includegraphics[width=1\linewidth]{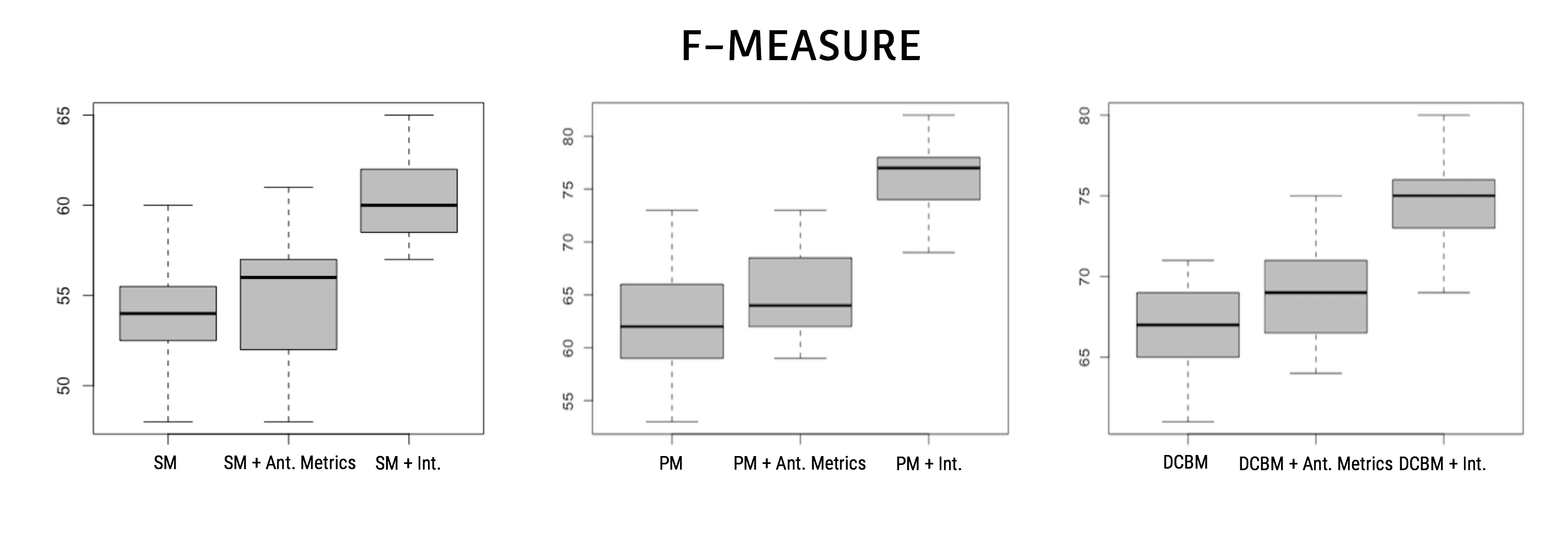}
	\caption{Overview of the value of F-measure among the models }\label{fig:fm}
\end{figure*}

\begin{figure*}[h]\centering
	\includegraphics[width=1\linewidth]{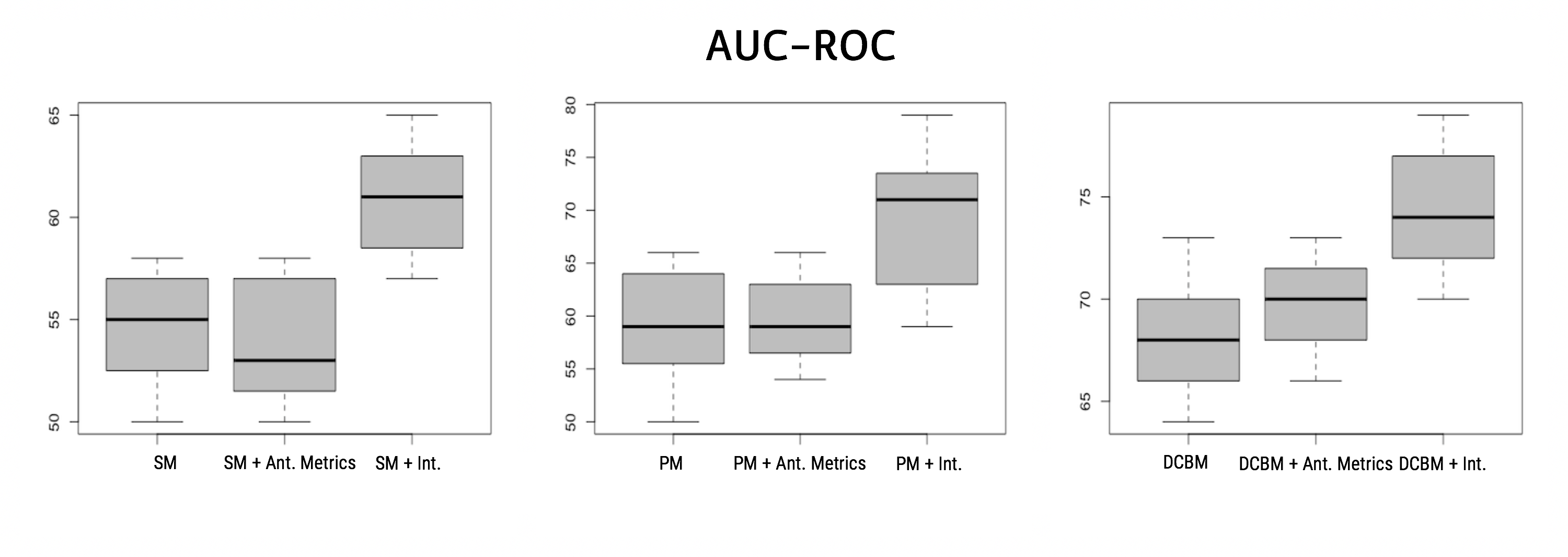}
	\caption{Overview of the value of AUC-ROC among the models }\label{fig:ar}
\end{figure*}

In this section we report and sum up the results of the presented research questions, discussing the main findings of our study.

\subsection{RQ$_1$-RQ$_2$: The performance of the intensity-including models and their comparison with the state of the art} 
Before describing the results related to the contribution of the intensity index in the three prediction models considered, we report the results of the feature selection process aimed at avoiding multi-collinearity. According to the results achieved using the \emph{vif} function \cite{Obrien2007}, we removed FCH, LCH, WFR, ATAF, CHD, LCD, CSBS, and ACDF from the process-based model \cite{elish2013suite}, while we did not remove any variables from the other baselines.

Figures \ref{fig:fm} and \ref{fig:ar} show the box plots reporting the distributions of F-Measure and AUC-ROC achieved by the (i) basic models that do not include any smell-related information - SM, PM, and DCBM, respectively; (ii) models including the antipattern metrics - those having ``+ Ant. Metrics'' as suffix; and (iii) models including the intensity index - those reporting ``+ Int.'' as suffix. 
Note that for the sake of readability, we only report the distribution of F-Measure rather than the distributions of precision and recall. Detailed results for those metrics are available in our online appendix \cite{appendix}.

In the first place, looking at Figure \ref{fig:fm}, we can observe that the basic model based on scattering metrics (\ie DCBM) tends to perform better than models built using structural and process metrics. Indeed, DCBM \cite{di2015role} has a median F-Measure 5\% and 13\% higher than structural (67\% vs 54\%) and process (67\% vs 62\%) models, respectively.
On the one hand, this result confirms our previous findings on the power of the developer-related factors in change prediction \cite{catolino2018enhancing}; on the other hand, we can confirm the results achieved by Di Nucci \etal \cite{di2015role} on the value of the scattering metrics for the prediction of problematic classes. 
As for the role of the intensity index, we notice that with the respect to the SM, PM and DCBM model, the intensity of code smells provides an additional useful information able to increase the ability of the model in discovering change-prone code components. This is observable by looking at the performance in Figures \ref{fig:fm} and \ref{fig:ar}. In the following, we further discuss our findings by reporting our results for each prediction model experimented, comparing the intensity-including ones with the state of the art. \medskip

\noindent \textbf{Contribution in Structural-based Models.} The addition of the intensity index within the SM model allows the model to reach a median F-Measure of 60\% and an AUC-ROC of 61\%, respectively. When compared against the antipattern metrics-including model, the intensity-including one still performs better (\ie +4\% in terms of median F-Measure and +7\% in terms of median AUC-ROC). 

Looking deeper into the results, we observed that the shapes of the box plots for the intensity-including model appear less dispersed than the basic one, meaning that the addition of the intensity index makes the performance of the model better and more stable. For instance, considering the \texttt{Apache-ant-1.3} project, the basic structural model reached 50\% precision and 56\% recall (F-Measure=53\%), while the model that includes the intensity index has a precision of 61\% and a recall of 66\% (F-Measure=63\%), thus obtaining an improvement of 10\%. The same happens all the considered systems: we can therefore claim that the performances of change prediction models strongly improve when considering the intensity of code smells as additional independent variable.

When considering the structural model that includes the antipattern metrics defined by Taba \etal \cite{Taba:2013}, we notice that its performance is just slightly better than the basic model in terms of F-Measure (56\% vs 54\%); more interesting, the AUC-ROC of the ``SM + Ant. Metrics'' model is lower than the basic one (53\% vs 55\%). From a practical perspective, these results tell us that the inclusion of the antipattern metrics only provides slight improvement with respect to the number of actual change-prone classes identified, but at the same time cannot provide benefits in the robustness of the classifications. 

In the comparison between the \emph{SM + Ant. Metrics} and the \emph{SM + Int.} models, we observe that the performance of the former is always lower than the one achieved by the latter (considering the median of the distributions, -4\% of F-Measure and -8\% of AUC-ROC). This indicates that the intensity index can provide much higher benefits in change prediction than existing metrics that capture other smell-related information. Nevertheless, in some cases the antipattern metrics defined by Taba \etal \cite{Taba:2013} can give complementary information with respect to the intensity index, opening the possibility to obtain still better performance by considering both metric types. Our claim is supported by the overlap analysis shown in Table \ref{tab:overlap} and computed on the set of \emph{change-prone} and \emph{smelly} classes correctly classified by the two models. While 43\% of the instances are correctly classified by both the models, a consistent portion of instances are classified only by {SM + Int.} model (35\%) or by the model using the antipattern metrics (22\%). Consequently, this means that the smell-related information taken into account by the \emph{SM + Int.} and \emph{SM + Ant. Metrics} models are orthogonal and complement each other.

The observations made above were also confirmed from a statistical point of view. Indeed, the intensity-including prediction model consistently appeared in the top Scott-Knott ESD rank in terms of AUC-ROC, meaning that its performance was statistically higher than the baselines in most of the cases (40 projects out of 43).\medskip

\noindent \textbf{Contribution in Process-based Models.} Also in this case the addition of the intensity index in the model defined by Elish \etal \cite{elish2013suite} improved its performance with respect to the basic model (PM). Indeed, the overall median value of F-Measure increased of 15\%, \ie F-Measure of \emph{PM + Int.} is 77\% while that of \emph{PM} is 62\%. An interesting aspect to discuss in this case is related to the ability of the intensity-including model to increase both precision and recall with respect to the basic model.
This is, for instance, the case of \texttt{Apache Ivy 2}, where PM reaches 61\% of precision and 49\% of recall; by adding the intensity index, the prediction model increases its performances to 76\% (+15\%) in terms of precision and 77\% (+28\%) of recall, demonstrating that a better characterization of the classes having design problems can help in obtaining more accurate predictions.

Looking at the baseline model that includes the antipattern metrics, we notice that it provides improvements when compared to the basic one. However, such improvements are still minor in terms of F-Measure (64\% vs 62\%) and thus we can confirm that the addition of the metrics proposed by Taba \etal \cite{Taba:2013} does not provide a relevant boost in the performance of basic change prediction models. Similarly, the model based on such metrics is never able to outperform the performance of the intensity-including one, being up to 13\% less performing. At the same time, it is worth reporting an interesting complementarity between the set of \emph{change-prone and smelly} classes correctly classified by \emph{``PM + Int.''} and by the \emph{Basic + Ant. Metrics} (see Table \ref{tab:overlap}), \ie the two models correctly capture the change-proneness of different code elements.

The statistical analyses confirmed the findings discussed above. Indeed, the likelihood to be ranked at the top by the Scott-Knott ESD test is always higher for the model including the intensity index. At the same time, the antipattern metrics-including model was confirmed to provide a slight statistical benefits than the basic one (they are ranked in the same cluster in 88\% of the cases). \medskip

\noindent \textbf{Contribution in Developer-Related Model.} Finally, the results for this type of model is similar to the one discussed above. Indeed, the addition of the intensity in DCBM \cite{di2015role} allows the model to reach a median F-Measure of 75\% and an AUC-ROC of 74\%, respectively. When compared to the standard model \emph{DCBM} the intensity-including one performs better (\ie +7\% in terms of median F-Measure and +6\% in terms of median AUC-ROC). For instance, in the \texttt{Apache Synapse 1.2} project the \emph{``DCBM + Int.''} obtains an F-Measure and AUC-ROC 12\% and 13\%, respectively, higher than \emph{DCBM}. The result holds for all the systems in our dataset, meaning that the addition of the intensity always provides improvements with respect to the baseline.

Comparing the performance of \emph{``DCBM + Int.''} with the model that includes the antipattern metrics \cite{Taba:2013}, we observe that the F-Measure of the former is on average 6\% higher than the latter; the better performance of the intensity-including model is also confirmed when considering the AUC-ROC, which is 4\% higher. Nevertheless, also in this case we found some complementarities in the correct predictions done by these two models (see Table \ref{tab:overlap}): indeed, only 44\% of instances are correctly caught by both the models, while 31\% of them are only captured by \emph{``DCBM + Int.''} and 25\% only by \emph{``DCBM + Ant. Metrics''}.
The Scott-Knott ESD test confirmed our observations. The likelihood of the intensity-including model to be ranked at the top is always higher than the other models. At the same time, the antipattern metrics-including models were confirmed to provide statistically better performance than the basic models in 67\% of the considered systems. \medskip

\begin{table}[h]
	\centering
	\caption{Overlap analysis between the model including the intensity index and the model including the antipattern metrics.}
	\label{tab:overlap}
	\resizebox{.7\linewidth}{!}{
		\begin{tabular}{lccc}
			\hline
			\multirow{2}{*}{ Models} & Int. $\cap$ & Int. $\setminus$ & Ant. $\setminus$\\
			& Ant.\% & Ant.\% & Int.\% \\\hline
			
			\rowcolor{gray!25} SM \cite{zhou2009examining} & 43 & 35 & 22 \\
			PM \cite{elish2013suite} & 47 & 38 & 15 \\
			\rowcolor{gray!25} DCBM \cite{di2015role} & 44 & 31 & 25 \\
		\end{tabular}
	}
	
\end{table}
\newpage
\begin{mybox}
	\textbf{RQ$_1$ - To what extent does the intensity index improve the performance of existing change prediction models?} The addition of the intensity index \cite{palomba2017toward} as a predictor of change-prone components increases the performance of the baseline change prediction models in terms of F-Measure up to 10\%. \medskip
\end{mybox}

\begin{mybox}
	\textbf{RQ$_2$ - How does the model including the intensity index as predictor compare to a model built using antipattern metrics?} The prediction models that include the antipattern metrics \cite{Taba:2013} only perform slightly better than the basic models, while they have lower performance than the intensity-including ones. However, we observed interesting complementarities between the set of \emph{change-prone and smelly} classes correctly classified by the models that include \emph{intensity index} and the models with \emph{antipattern metrics}, which highlight the possibility to achieve higher performance through a combination of smell-related information.
\end{mybox}

\subsection{RQ$_3$: Analyzing the gain provided by the intensity index with respect to other predictors} 
In this section we analyze the results of \emph{Gain Ratio Feature Evaluation} algorithm \cite{Quinlan:1986} in order to understand how important the predictors composing the different models considered in this study are, with the aim to evaluate the predictive power of the intensity index when compared to the other predictors. 

Table \ref{tab:gainSM} shows the gain provided by the different predictions employed in the structural metrics-based change prediction model, while Table \ref{tab:gainPM} reports the results for the process-based model and Table \ref{tab:gainDCBM} those for the DCBM model. In particular, the tables report the ranking of the predictors based on their importance within the individual models through the values of the mean and the standard deviation (computed by considering the results obtained on the single systems) of the expected reduction in entropy caused by partitioning the prediction model according to a given predictor. In addition, we also provide the likelihood of the predictor to be in the top-rank by the Scott-Knott ESD test, \ie the percentage of times a predictor was statistically better than the others. The following subsections discuss our findings considering each prediction model individually. \medskip

\begin{table}[h]
	\centering
	\caption{Gain Provided by Each Metric To The SM Prediction Model.}
	\label{tab:gainSM}
	\resizebox{.7\linewidth}{!}{
		\begin{tabular}{lccc}
			\toprule
			\multirow{2}{*}{Metric} & \multirow{2}{*}{Mean} & \multirow{2}{*}{St. Dev.}& SK-ESD \\
			& & &Likelihood \\\hline
			CBO & 0.66 & 0.09 & 82 \\
			\rowcolor{gray!25} RFC & 0.61 & 0.05 & 77\\
			\textbf{Intensity} & \textbf{0.49} & \textbf{0.13} & \textbf{75} \\
			\rowcolor{gray!25} LOC & 0.44 & 0.11 & 55 \\
			LCOM & 0.43 & 0.12 & 51\\
			\rowcolor{gray!25} Antipattern Complexity Metric & 0.42 & 0.12 & 41 \\
			Antipattern Recurrence Length & 0.31 & 0.05 & 32 \\
			\rowcolor{gray!25} Average Number of Antipatterns & 0.22 & 0.10 & 21 \\
			DIT & 0.13 & 0.02 & 3\\\hline
	\end{tabular} }
	
\end{table}

\noindent \textbf{Gain Provided to Structural-based Models \cite{zhou2009examining}.} The results in Table \ref{tab:gainSM} shows how Coupling Between Objects (CBO) is the metric having the highest predictive power, with an average reduction of entropy of 0.66 and a standard deviation of 0.09. The Scott-Knott ESD test statistically confirmed the importance of the predictor, since the information gain given by the metric was statistically higher than other metrics in 82\% of the cases. It is worth noting that this result is in line with previous findings in the field \cite{basili:1996,palomba2017toward,zhou2009examining} which showed the relevance of coupling information for the maintainability of software classes. Looking at the ranking, we also noticed that Response For a Class (RFC), Lines of Code (LOC), and Lack of Cohesion of Methods (LCOM3) appear to be relevant. On the one hand, this is still in line with previous findings \cite{basili:1996,palomba2017toward,zhou2009examining}. On the other hand, it is also important to note that our results seem to reconsider the role of code size for assessing change prediction. In particular, unlike the findings by Zhou \etal \cite{zhou2009examining} on the confounding effect of size, we discovered that LOC can be an important predictor to discriminate change-prone classes. This may be due to the large dataset exploited in this study, which allows a higher level of generalizability. The Scott-Knott ESD test confirmed that these metrics are among the most powerful ones.

As for the variable of interest, \ie the intensity index, we could observe that it is the feature providing the third highest gain in terms of reduction of entropy, as it has a value of Mean and Standard Deviation of 0.49 and 0.13, respectively. Looking at the results of the statistical test, we observed that the intensity index is ranked on the top by the Scott-Knott ESD in 49\% of the cases, thus confirming the high predictive power of the metric. These findings lead to two main observations. In the first place, the intensity index is a relevant variable and for this reason can provide high benefits for the prediction of change-prone classes (as also observed in \textbf{RQ$_1$}). Secondly, and perhaps more interesting, the intensity index can be more powerful than other structural metrics from which it is derived: in other words, a metric mixing together different structural aspects to measure how severe a code smell is seems to be more meaningful than the individual metrics used to derive the index.

As for the antipattern metrics, we observed that all of them appear to be less relevant than the intensity index. Somehow, this confirms the results of \textbf{RQ$_2$}, where we showed that adding them to change prediction models results in a limited improvement with respect to the baseline. At the same time, it is worth noting that ACM (\ie \emph{Antipattern Complexity Metric}) may sometimes provide a notable contribution. While the average gain is 0.42, the standard deviation is 0.12: this means that the entropy reduction can be up to 0.54, as in the case of the \texttt{Apache-synapse-2.3}. Thus, this result seems to suggest that this metric has some potential for effectively predicting the change-proneness of classes. The other two antipattern metrics, \ie \emph{Average Number of Antipatterns} (ANA) and \emph{Antipattern Recurrence Length} (ARL), instead, only provide a partial contribution in the reduction of entropy of the model. Indeed, the mean are 0.31 and 0.22, respectively, with a standard deviation that is never above 0.1. The Scott-Knott ESD test statistically confirmed the findings: indeed, ACM was a top predictor in 41\% of the datasets, as opposed to ANA and ARL metrics which appeared as statistically more powerful than other metrics only in 32\% and 21\% of the cases.
Finally, Depth Inheritance Tree was the less powerful metrics in the ranking, and the Scott-Knott ESD test ranked it at the top in only 3\% of the cases. \medskip

\begin{table}[h]
	\centering
	\caption{Gain Provided by Each Metric To The PM Prediction Model.}
	\label{tab:gainPM}
	\resizebox{.7\linewidth}{!}{
		\begin{tabular}{lccc}
			\toprule
			\multirow{2}{*}{Metric} & \multirow{2}{*}{Mean} & \multirow{2}{*}{St. Dev.}& SK-ESD \\
			& & &Likelihood \\\hline
			BOC & 0.56 & 0.05 & 75 \\
			\rowcolor{gray!25} FRCH & 0.55 & 0.06 & 64\\
			\textbf{Intensity} & \textbf{0.44} & \textbf{0.08} & \textbf{61} \\
			\rowcolor{gray!25} WCD & 0.42 & 0.11 & 55\\
			Antipattern Complexity Metric & 0.41 & 0.04 & 54 \\
			\rowcolor{gray!25} LCA & 0.33 & 0.07 & 33 \\
			CHO & 0.28 & 0.03 & 31 \\
			\rowcolor{gray!25} Antipattern Recurrence Length & 0.24 & 0.05 & 25 \\
			Average Number of Antipatterns & 0.09 & 0.03 & 2 \\
			\rowcolor{gray!25} CSB & 0.07 & 0.01 &1 \\
			TACH& 0.02 & 0.01 & 1 \\\hline
	\end{tabular} }
	
\end{table}

\noindent \textbf{Gain Provided to Process-based Models \cite{elish2013suite}.} Regarding the \textit{process metric-based} model considered in this study, the results are similar to the structural model. Indeed, from Table \ref{tab:gainPM} it is evident that the intensity index represents a good predictor; the value of the mean is 0.44 and it is a top predictors in 61\% of the dataset. It appears to be the third most powerful feature of the model, just behind the Birth of a Class and Frequency of Changes. On the one hand, this ranking is pretty expected, as the top two features are those which fundamentally characterize the notion of process-based change prediction proposed by Elish \etal \cite{elish2013suite}. On the other hand, our findings report that the intensity index can effectively complement the process metrics present in the model, \ie a structural-based indicator seems to be orthogonal with respect to the other basic features. As for the antipattern metrics, also in this case ACM turned to be a potentially good predictor in 54\% of the dataset, while ANA and ARL are top predictors only in 25\% and 2\% of the dataset, respectively. At the bottom of the ranking there are other basic metrics like Changes since the Birth and Total Amount of Changes: this confirms previous findings \cite{catolino2018enhancing} reporting that the overall number of previous changes cannot properly model the change-proneness of classes.

\begin{table}[h]
	\centering
	\caption{Gain Provided by Each Metric To The DCBM Prediction Model.}
	\label{tab:gainDCBM}
	\resizebox{.7\linewidth}{!}{
		\begin{tabular}{lccc}
			\toprule
			\multirow{2}{*}{Metric} & \multirow{2}{*}{Mean} & \multirow{2}{*}{St. Dev.}& SK-ESD \\
			& & &Likelihood \\\hline
			Semantic Scattering & 0.76 & 0.07 & 95 \\
			\rowcolor{gray!25} \textbf{Intensity} & \textbf{0.74} & \textbf{0.05} & \textbf{94} \\
			Structural Scattering & 0.72 & 0.05 & 91\\
			\rowcolor{gray!25} Antipattern Complexity Metric & 0.66 & 0.04 & 78 \\
			Antipattern Recurrence Length & 0.31 & 0.02 & 44 \\
			\rowcolor{gray!25} Average Number of Antipatterns & 0.11 & 0.03 & 21 \\\hline
	\end{tabular} }
	
\end{table}

\noindent \textbf{Gain Provided to Developer-related factors \cite{di2015role}.}  Looking at the ranking of the features of the DCBM model, we can still confirm the results discussed so far. Indeed, the intensity index is the second most relevant factor, just behind the semantic scattering: its mean is 0.74, and the Scott-Knott ESD test indicated the intensity index as top predictor in 94\% of the dataset. It is interesting to note that the index provides a higher contribution than the structural scattering, meaning that the combination from which it is derived can provide a higher entropy reduction with respect to a structural metric that computes how far the classes touched by developers in a certain time window are.  Regarding the antipattern metrics, the results are similar to those of the other models considered; the ACM provided a pretty high quantity of additional information to the model (mean=0.66), being ranked at top predictors in 78\% of the dataset. Instead, the means for ARL and ANA are notably lower (0.31 and 0.11, respectively), appearing as the least important features.

As a more general observation, it is worth noting that the values of mean information gain of both the intensity index and ACM are much higher ($\approx$0.20 more) for this model when compared to the structural- and process metrics-based models. This indicates that those metrics can provide a much higher information than the other models: this can be due to the limited number of features employed by this model, which makes the additional metrics more useful to predict the change-proneness of classes.

All in all, we can confirm again that the intensity index provides a strong information gain to all the change prediction models considered in the study, together with ACM from the group of antipattern metrics. This possibly highlights how their combination could provide further improvements in the context of change prediction.\medskip

\begin{mybox}
	\textbf{RQ$_3$ - What is the gain provided by the intensity index to change prediction models when compared to other predictors?} The intensity index provides a notable information gain in all the considered prediction models. At the same time, a metric of complexity of the change process involving code smells seems to provide further additional information, highlighting the possibility to obtain even better change prediction performance when mixing different smell-related information.
\end{mybox}

\subsection{RQ$_4$: The performance of the combined smell-aware bug prediction model}
The results achieved in the previous research questions highlight the possibility to build a combined change prediction model that takes into account smell-related information besides the structural, process, and developer-related metrics. For this reason, in the context of \textbf{RQ$_4$}, we assessed the feasibility of a combined solution and evaluated its performance with respect to the results achieved by the stand-alone models that only rely on structural, process, or developer-related features.
As explained in Section \ref{sec:study}, to come up with the combined model we firstly put together all the features of the considered models and then applied a feature selection algorithm to discard irrelevant features. Starting from an initial set of 18 metrics, this procedure discarded DIT, CSB, TACH, and ANA. Thus, the combined model comprises 14 metrics: besides most of the basic structural, process, and developer-related predictors, the model includes three smell-related metrics such as (i) intensity index, (ii) ACM, and (iii) ARL. 

\begin{figure*}[h]\centering
	\includegraphics[width=\linewidth]{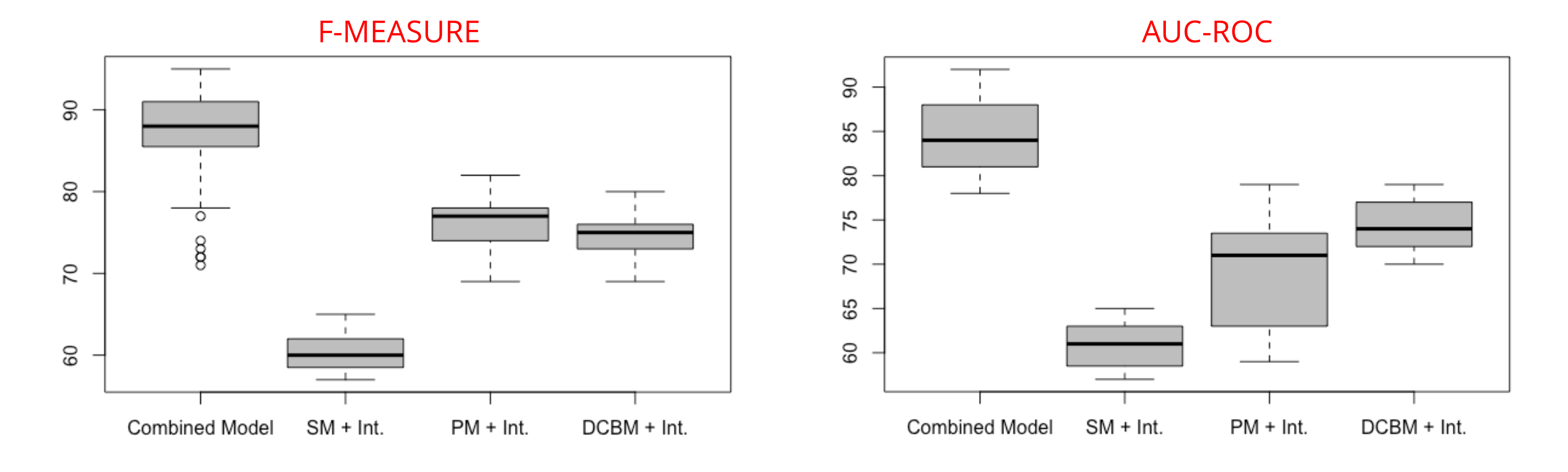}
	\caption{Overview of the value of F-Measure and AUC-ROC of the Combined Model}\label{fig:combined}
\end{figure*}

Figure \ref{fig:combined} shows the boxplots reporting the distributions of F-Measure and AUC-ROC related to the smell-aware combined change prediction model. To facilitate the comparison with the models exploited in the context of \textbf{RQ$_1$} and \textbf{RQ$_2$}, we also report boxplots depicting the best models coming from our previous analyses, \ie \emph{SM + Int.}, \emph{PM + Int.}, and \emph{DCBM + Int.}

Looking at the figures, it seems clear that the combined model has better performances than all the baseline approaches that exploit individual features. Indeed, the median F-Measure reaches 88\%, meaning that it is 18\%, 11\%, and 13\% more accurate than \emph{SM + Int.}, \emph{PM + Int.}, and \emph{DCBM + Int.}, respectively. These results hold when considering the AUC-ROC, where the combined model was able to boost the performance at least 10\% with respect to the basic models that include the intensity.
As an example, in the \texttt{Apache Xalan 2.5} project the best stand-alone model (the \emph{``SM + Int''} in this case) had an F-Measure close to 73\%, while the mixture of features provided by the combined model allowed to reach an F-Measure of 93\%. As expected, the results are statistically significant, as the devised smell-aware change prediction model appeared in top Scott-Knott ESD rank in 98\% of the cases. 

On the one hand, these strong results confirm previous findings on the importance to combine different predictors of source code maintainability \cite{catolino2018enhancing,DAmbros:emse}. On the other hand, we can claim that smell-related information \cite{palomba2017toward,Taba:2013} improves the capabilities of change prediction models, allowing them to perform much better than other existing models.\medskip
\begin{mybox}
	\textbf{RQ$_4$ - What is the performance of a combined change prediction model that includes smell-related information?} The devised smell-aware change prediction model performs better than all the baseline approaches considered in the paper, with an F-Measure up to 20\%.
\end{mybox}

\section{Threats to Validity}
\label{sec:threats}
	
In this section we discuss possible threats affecting our results and how we mitigated them.

\subsection{Threats to Construct Validity}
As for threats related to the relationship between theory and observation, they are mainly related to the independent variables used and the dataset exploited. As for the former, we selected state of the art change prediction models based on a different set of basic features, \ie structural, process, and developer-related metrics, that capture different characteristics of source code. The selection was mainly driven by recent results \cite{catolino2018enhancing} that showed that the considered models are (i) accurate in the detection of the change-proneness of classes and (ii) complementary to each other, thus correctly identifying different sets of change-prone classes. All in all, this selection process allowed us to test the contribution of smell-related information in different contexts. As for the dataset, we relied on a publicly available source previously built \cite{palomba2017toward}. Of course, we cannot exclude possible imprecisions in the computation of the dependent variable. 
Still in this category, we adopted \textsc{JCodeOdor} \cite{fontana2015towards} to identify code smells and assign to them a level of intensity. Our choice was driven by previous results \cite{palomba2017toward} that showed, on the same dataset, that this tool has a high accuracy (\ie F-Measure=80\%). Despite this performance, the tool still identifies 154 false positives and 94 false negatives among the 43 considered systems. To make the set of code smells as close as possible to the \emph{ground truth}, in our study we manually elaborated on the output of the tool by (i) setting to zero the intensity index of the false positive instances, and (ii) discarding the false negatives, \ie the instances for which we could not assign an intensity value. Since this manual process is not always feasible, we also evaluated the effect of including false positive and false negative instances in the construction of the change prediction models. 
More specifically, we re-ran the analyses performed in Section \ref{sec:study} and validated the performance of the experimented models when including the false positive instances using the same metrics used to assess the performance of the other prediction models (\ie F-Measure and AUC-ROC). Our results report that these models always perform better than other models that do not include any smell-related information, while they are slightly less performing (-3\% in terms of median F-Measure) than those built discarding the false positive instances.

In the second place, we evaluated what is the impact of including false negative instances. Their intensity index is, by definition, equal to zero: as a consequence, they were considered in the same way as \emph{non-smelly} classes. The results of our analyses showed that the \emph{intensity-including} models still produce better results than those of the baselines, as they boosted their median F-Measure of $\approx$4\%. At the same time, we observed a decrement of 2\% in terms of F-Measure with respect to the performance obtained by the prediction models built discarding false negatives.

As a final step, we also considered the case where both false positives and false negatives are incorporated in the experimented models. Our findings report that the \emph{Basic + Intensity} models have a median F-Measure 2\% lower than the models where the false positive and false negative instances were filtered out. At the same time, they were still better than basic models (median F-Measure=+6\%). Thus, we could conclude that \emph{a fully automatic code smell detection process still provides better performance than existing change prediction models.} This result is, in our opinion, extremely valuable as it indicates that practitioners can adopt automatic code smell detectors without the need of manually investigating the candidates they give as output. 

It is worth remarking that the choice of considering code smell severity rather than the simple presence/absence of smells was driven by the conjecture that the severity can give a more fine-grained information on how much a design problem is harmful for a certain source code class. To verify this conjecture, we conducted a further analysis aimed at establishing the performance of the experimented models where considering a boolean value reporting the presence of code smells rather than their intensity. As expected, our findings reported that the models relying on the intensity were more powerful than those based on the boolean indication of the smell presence. This further confirms the idea behind this paper, \ie code smell intensity can improve change-proneness prediction.

Our observations might still have been threatened by the presence of a large number of code smell co-occurrences \cite{palomba2018large,Yamashita:ICSE2013}, which might have biased the intensity level of the smelly classes of our dataset. To account for this aspect, we measured the percentage of classes in our dataset affected by more than one smell: we only found that 8\% of the classes, on average, contained more code smells. As a consequence, we can claim that the problem of co-occurrence is rather limited in our study.

A further threat to construct validity relates to the dataset exploited in our empirical study. In this regard, we rely on a publicly available oracles from the PROMISE repository \cite{menzies2012promise}, however we performed some preliminary operations to ensure data quality and robustness, by employing the data preprocessing algorithm implementing the guidelines by Shepperd \etal \cite{Shepperd} aimed at removing noisy and/or erroneous data items.
 
\subsection{Threats to Conclusion Validity}
Threats to \emph{conclusion validity} refer to the relation between treatment and outcome. 
When evaluating the change prediction models we computed well-established metrics such as F-Measure and AUC-ROC. Furthermore, we statistically verified the differences in the performance achieved by the different experimented models using the Scott-Knott statistical test \cite{ScottKnott} and Cliff's Delta~\cite{cliff} for measuring the effect size. Moreover, we analyzed to what extent the intensity index is important with respect to the other metrics by analyzing the gain provided by the addition of the severity measure in the model. Finally, to ensure that the experimented models did not suffer from multi-collinearity, we adopted the \textit{variance inflation factors} function \cite{Obrien2007} to discard non-relevant variables from the considered features.

%In the context of \textbf{RQ3}, we measured the contribution of the intensity index in bug prediction model performances by exploiting the \emph{Gain Ratio Feature Evaluation} algorithm \cite{Quinlan:1986}. While the usage of other procedures (\eg the Wrapper approach \cite{Kohavi:1997}) might have lead to different results, it is important to note that this algorithm was the most suitable for our study since it allowed us to quantify the exact entropy reduction achieved using the intensity index.

Still in this category there is a possible threat related to the validation methodology exploited. As shown by Tantithamthavorn \etal \cite{Tantithamthavorn:2017}, ten-fold cross validation might provide unstable results because of the effect of random splitting. To deal with this issue, we repeated the 10-fold cross validation 100 times: in this way, we drastically removed the bias due to the validation strategy.

\subsection{Threats to External Validity}
Threats in this category mainly concern the generalization of results. In this case we analyzed a large set of 43 releases of 14 software systems coming from different application domains and with different characteristics (size, number of classes, etc.). Another threat in this category regards the choice of the baseline models. However, we evaluated the contribution of the smell-related information in the context of change prediction models widely used in the past \cite{catolino2018enhancing,elish2013suite,zhou2009examining} that take into account predictors of different nature, \ie, product, process, and developer-related metrics. However, we are aware that our study is based on systems developed in Java only, and therefore future investigations aimed at corroborating our findings on a different set of systems would be worthwhile.

\section{Conclusion}
\label{sec:conclusion}

Change-prone classes represent code components that tend to change more often, because of their importance for the business logic of a system or because not properly designed by developers. To help practitioners in keeping maintenance and evolution activities under control, many change prediction models have been defined in literature \cite{catolino2018enhancing,elish2013suite,zhou2009examining}. Following the findings of previous studies \cite{khomh2012exploratory,palomba2017diffuseness} reporting the impact of code smells on the change-proneness of classes, in this paper we aimed at investigating the impact of smell-related information for the prediction of change-prone classes. We first conducted a large empirical study on 43 releases of 14 software systems and evaluated the contribution of the intensity index proposed by Arcelli Fontana \etal \cite{ArcelliFontana2015h} within existing change prediction models based on structural- \cite{zhou2009examining}, process-\cite{elish2013suite}, and developer-related \cite{di2015role} metrics. We also compared the gain provided by the intensity index with the one given by the so-called antipattern metrics \cite{Taba:2013}, \ie metrics capturing historical aspects of code smells.

The results indicated how the addition of the intensity index as a predictor of change-prone components increases the performance of baseline change prediction models by an average of 10\% in terms of F-Measure (RQ$_1$). Moreover, the intensity index can boost the performance of such models more than state of the art smell-related metrics such as those defined by Taba \etal \cite{Taba:2013}, even though we observed some complementarities between the models exploiting different information on code smells(RQ$_2$). Based on these results, we built a combined smell-aware change prediction model that takes into account product, process, developer- and smell-related information(RQ$_4$). The key results showed how the combined model provides a consistent boost in terms of F-Measure, which goes up to 20\% more.

Our findings represent the main input for our future research agenda: we first aim at further testing the usefulness of the devised model in an industrial setting. Furthermore, we plan to perform a fine-grained analysis into the role of each smell type independently on the change prediction power.

\bibliographystyle{spbasic}
\bibliography{main}

\begin{thebibliography}{128}
\providecommand{\natexlab}[1]{#1}
\providecommand{\url}[1]{{#1}}
\providecommand{\urlprefix}{URL }
\expandafter\ifx\csname urlstyle\endcsname\relax
  \providecommand{\doi}[1]{DOI~\discretionary{}{}{}#1}\else
  \providecommand{\doi}{DOI~\discretionary{}{}{}\begingroup
  \urlstyle{rm}\Url}\fi
\providecommand{\eprint}[2][]{\url{#2}}

\bibitem[{Abbes et~al(2011)Abbes, Khomh, Gueheneuc, and
  Antoniol}]{abbes2011empirical}
Abbes M, Khomh F, Gueheneuc YG, Antoniol G (2011) An empirical study of the
  impact of two antipatterns, blob and spaghetti code, on program
  comprehension. In: Software maintenance and reengineering (CSMR), 2011 15th
  European conference on, IEEE, pp 181--190

\bibitem[{Abdi et~al(2006)Abdi, Lounis, and Sahraoui}]{abdi2006analyzing}
Abdi M, Lounis H, Sahraoui H (2006) Analyzing change impact in object-oriented
  systems. In: 32nd EUROMICRO Conference on Software Engineering and Advanced
  Applications (EUROMICRO'06), IEEE, pp 310--319

\bibitem[{Aniche et~al(2016)Aniche, Treude, Zaidman, van Deursen, and
  Gerosa}]{Aniche:scam16}
Aniche M, Treude C, Zaidman A, van Deursen A, Gerosa M (2016) Satt: Tailoring
  code metric thresholds for different software architectures. In: 2016 IEEE
  16th IEEE International Working Conference on Source Code Analysis and
  Manipulation (SCAM)

\bibitem[{Arcelli~Fontana et~al(2015)Arcelli~Fontana, Ferme, Zanoni, and
  Roveda}]{ArcelliFontana2015h}
Arcelli~Fontana F, Ferme V, Zanoni M, Roveda R (2015) Towards a prioritization
  of code debt: A code smell intensity index. In: Proceedings of the Seventh
  International Workshop on Managing Technical Debt (MTD 2015), IEEE, Bremen,
  Germany, pp 16--24, in conjunction with ICSME 2015

\bibitem[{Arcelli~Fontana et~al(2016)Arcelli~Fontana, M{\"a}ntyl{\"a}, Zanoni,
  and Marino}]{ArcelliFontana:emse2016}
Arcelli~Fontana F, M{\"a}ntyl{\"a} MV, Zanoni M, Marino A (2016) Comparing and
  experimenting machine learning techniques for code smell detection. Empirical
  Software Engineering 21(3):1143--1191, \doi{10.1007/s10664-015-9378-4},
  \urlprefix\url{http://dx.doi.org/10.1007/s10664-015-9378-4}

\bibitem[{Arisholm et~al(2004)Arisholm, Briand, and
  Foyen}]{arisholm2004dynamic}
Arisholm E, Briand LC, Foyen A (2004) Dynamic coupling measurement for
  object-oriented software. IEEE Transactions on Software Engineering
  30(8):491--506

\bibitem[{Bacchelli and Bird(2013)}]{bacchelli:2013}
Bacchelli A, Bird C (2013) Expectations, outcomes, and challenges of modern
  code review. In: Proc. of the International Conference on Software
  Engineering (ICSE), IEEE, pp 712--721

\bibitem[{Baeza-Yates and Ribeiro-Neto(1999{\natexlab{a}})}]{Baeza-Yates:1999}
Baeza-Yates R, Ribeiro-Neto B (1999{\natexlab{a}}) Modern Information
  Retrieval. Addison-Wesley

\bibitem[{Baeza-Yates et~al(1999)Baeza-Yates, Ribeiro-Neto
  et~al}]{baeza1999modern}
Baeza-Yates R, Ribeiro-Neto B, et~al (1999) Modern information retrieval, vol
  463. ACM press New York

\bibitem[{Baeza-Yates and Ribeiro-Neto(1999{\natexlab{b}})}]{BaezaYates}
Baeza-Yates RA, Ribeiro-Neto B (1999{\natexlab{b}}) Modern Information
  Retrieval. Addison-Wesley Longman Publishing Co., Inc.

\bibitem[{Bansiya and Davis(2002)}]{bansiya:2002}
Bansiya J, Davis CG (2002) A hierarchical model for object-oriented design
  quality assessment. IEEE Trans Softw Eng 28(1):4--17

\bibitem[{Basili et~al(1996)Basili, Briand, and Melo}]{basili:1996}
Basili VR, Briand LC, Melo WL (1996) A validation of object-oriented design
  metrics as quality indicators. IEEE Trans Softw Eng 22(10):751--761

\bibitem[{Bell et~al(2013)Bell, Ostrand, and Weyuker}]{bell2013limited}
Bell RM, Ostrand TJ, Weyuker EJ (2013) The limited impact of individual
  developer data on software defect prediction. Empirical Softw Engg
  18(3):478--505

\bibitem[{Bieman et~al(2003)Bieman, Straw, Wang, Munger, and
  Alexander}]{bieman:metrics}
Bieman JM, Straw G, Wang H, Munger PW, Alexander RT (2003) Design patterns and
  change proneness: an examination of five evolving systems. In: Proc. Int'l
  Workshop on Enterprise Networking and Computing in Healthcare Industry, pp
  40--49, \doi{10.1109/METRIC.2003.1232454}

\bibitem[{Bottou and Vapnik(1992)}]{bottou:1992}
Bottou L, Vapnik V (1992) Local learning algorithms. Neural Comput
  4(6):888--900

\bibitem[{Boussaa et~al(2013)Boussaa, Kessentini, Kessentini, Bechikh, and
  Ben~Chikha}]{boussaa:2013}
Boussaa M, Kessentini W, Kessentini M, Bechikh S, Ben~Chikha S (2013)
  Competitive coevolutionary code-smells detection. In: Search Based Software
  Engineering, Lecture Notes in Computer Science, vol 8084, Springer Berlin
  Heidelberg, pp 50--65

\bibitem[{Briand et~al(1999)Briand, Wust, and Lounis}]{briand1999using}
Briand LC, Wust J, Lounis H (1999) Using coupling measurement for impact
  analysis in object-oriented systems. In: Proc. Int'l Conf. on Software
  Maintenance (ICSM), IEEE, pp 475--482

\bibitem[{Catolino and Ferrucci(2018)}]{catolino2018m}
Catolino G, Ferrucci F (2018) Ensemble techniques for software change
  prediction: A preliminary investigation. In: Machine Learning Techniques for
  Software Quality Evaluation (MaLTeSQuE), 2018 IEEE Workshop on, IEEE, pp
  25--30

\bibitem[{Catolino et~al(2018{\natexlab{a}})Catolino, Palomba, Arcelli~Fontana,
  De~Lucia, Ferrucci, and Zaidman}]{appendix}
Catolino G, Palomba F, Arcelli~Fontana F, De~Lucia A, Ferrucci F, Zaidman A
  (2018{\natexlab{a}}) Improving change prediction models with code
  smell-related information - replication package -
  \url{https://figshare.com/s/f536bb37f3790914a32a}

\bibitem[{Catolino et~al(2018{\natexlab{b}})Catolino, Palomba, De~Lucia,
  Ferrucci, and Zaidman}]{catolino2018enhancing}
Catolino G, Palomba F, De~Lucia A, Ferrucci F, Zaidman A (2018{\natexlab{b}})
  Enhancing change prediction models using developer-related factors. Journal
  of Systems and Software

\bibitem[{le~Cessie and van Houwelingen(1992)}]{leCessie:1992}
le~Cessie S, van Houwelingen J (1992) Ridge estimators in logistic regression.
  Applied Statistics 41(1):191--201

\bibitem[{D'Ambros et~al(2010)D'Ambros, Bacchelli, and Lanza}]{dambros:qsic}
D'Ambros M, Bacchelli A, Lanza M (2010) On the impact of design flaws on
  software defects. In: Quality Software (QSIC), 2010 10th International
  Conference on, pp 23--31, \doi{10.1109/QSIC.2010.58}

\bibitem[{DAmbros et~al(2012)DAmbros, Lanza, and Robbes}]{DAmbros:emse}
DAmbros M, Lanza M, Robbes R (2012) Evaluating defect prediction approaches: a
  benchmark and an extensive comparison. Empirical Software Engineering
  17(4):531--577

\bibitem[{Di~Nucci et~al(2017)Di~Nucci, Palomba, {De Rosa}, Bavota, Oliveto,
  and {De Lucia}}]{di2015role}
Di~Nucci D, Palomba F, {De Rosa} G, Bavota G, Oliveto R, {De Lucia} A (2017) A
  developer centered bug prediction model. IEEE Trans on Softw Engineering p to
  appear

\bibitem[{Di~Nucci et~al(2018)Di~Nucci, Palomba, Tamburri, Serebrenik, and
  De~Lucia}]{di2018detecting}
Di~Nucci D, Palomba F, Tamburri DA, Serebrenik A, De~Lucia A (2018) Detecting
  code smells using machine learning techniques: are we there yet? In: 2018
  IEEE 25th International Conference on Software Analysis, Evolution and
  Reengineering (SANER), IEEE, pp 612--621

\bibitem[{{Di Penta} et~al(2008){Di Penta}, Cerulo, Gueheneuc, and
  Antoniol}]{dipenta:icsm2008}
{Di Penta} M, Cerulo L, Gueheneuc YG, Antoniol G (2008) An empirical study of
  the relationships between design pattern roles and class change proneness.
  In: Proc. Int'l Conf. on Software Maintenance (ICSM), IEEE, pp 217--226,
  \doi{10.1109/ICSM.2008.4658070}

\bibitem[{Elish and Al-Rahman Al-Khiaty(2013)}]{elish2013suite}
Elish MO, Al-Rahman Al-Khiaty M (2013) A suite of metrics for quantifying
  historical changes to predict future change-prone classes in object-oriented
  software. Journal of Software: Evolution and Process 25(5):407--437

\bibitem[{Eski and Buzluca(2011)}]{eski2011empirical}
Eski S, Buzluca F (2011) An empirical study on object-oriented metrics and
  software evolution in order to reduce testing costs by predicting
  change-prone classes. In: Proc. Int'l Conf Software Testing, Verification and
  Validation Workshops (ICSTW), IEEE, pp 566--571

\bibitem[{Fluri et~al(2007)Fluri, Wuersch, PInzger, and Gall}]{fluri2007change}
Fluri B, Wuersch M, PInzger M, Gall H (2007) Change distilling: Tree
  differencing for fine-grained source code change extraction. IEEE
  Transactions on software engineering 33(11)

\bibitem[{Fontana and Zanoni(2017)}]{fontana2017code}
Fontana FA, Zanoni M (2017) Code smell severity classification using machine
  learning techniques. Knowledge-Based Systems 128:43--58

\bibitem[{Fontana et~al(2013)Fontana, Zanoni, Marino, and
  Mantyla}]{Fontana:icsm}
Fontana FA, Zanoni M, Marino A, Mantyla MV (2013) Code smell detection: Towards
  a machine learning-based approach. In: Software Maintenance (ICSM), 2013 29th
  IEEE International Conference on, pp 396--399, \doi{10.1109/ICSM.2013.56}

\bibitem[{Fontana et~al(2015{\natexlab{a}})Fontana, Ferme, Zanoni, and
  Roveda}]{fontana2015towards}
Fontana FA, Ferme V, Zanoni M, Roveda R (2015{\natexlab{a}}) Towards a
  prioritization of code debt: A code smell intensity index. In: 2015 IEEE 7th
  International Workshop on Managing Technical Debt (MTD), IEEE, pp 16--24

\bibitem[{Fontana et~al(2015{\natexlab{b}})Fontana, Ferme, Zanoni, and
  Yamashita}]{fontana2015automatic}
Fontana FA, Ferme V, Zanoni M, Yamashita A (2015{\natexlab{b}}) Automatic
  metric thresholds derivation for code smell detection. In: Proceedings of the
  Sixth international workshop on emerging trends in software metrics, IEEE
  Press, pp 44--53

\bibitem[{Fontana et~al(2016)Fontana, Dietrich, Walter, Yamashita, and
  Zanoni}]{Fontana:saner16}
Fontana FA, Dietrich J, Walter B, Yamashita A, Zanoni M (2016) Antipattern and
  code smell false positives: Preliminary conceptualization and classification.
  In: 2016 IEEE 23rd International Conference on Software Analysis, Evolution,
  and Reengineering (SANER), vol~1, pp 609--613, \doi{10.1109/SANER.2016.84}

\bibitem[{Fowler et~al(1999)Fowler, Beck, Brant, Opdyke, and
  Roberts}]{fowler:1999}
Fowler M, Beck K, Brant J, Opdyke W, Roberts D (1999) Refactoring: Improving
  the Design of Existing Code. Addison-Wesley

\bibitem[{Gatrell and Counsell(2015)}]{Gatrell:jss}
Gatrell M, Counsell S (2015) The effect of refactoring on change and
  fault-proneness in commercial c\# software. Science of Computer Programming
  102(0):44 -- 56, \doi{http://dx.doi.org/10.1016/j.scico.2014.12.002},
  \urlprefix\url{http://www.sciencedirect.com/science/article/pii/S0167642314005711}

\bibitem[{Ghotra et~al(2015)Ghotra, McIntosh, and Hassan}]{Ghotra:icse15}
Ghotra B, McIntosh S, Hassan AE (2015) Revisiting the impact of classification
  techniques on the performance of defect prediction models. In: Proceedings of
  the International Conference on Software Engineering, IEEE, pp 789--800

\bibitem[{Girba et~al(2004)Girba, Ducasse, and Lanza}]{girba2004yesterday}
Girba T, Ducasse S, Lanza M (2004) Yesterday's weather: Guiding early reverse
  engineering efforts by summarizing the evolution of changes. In: Proc. Int'l
  Conf. Softw. Maintenance (ICSM), IEEE, pp 40--49

\bibitem[{Grissom and Kim(2005{\natexlab{a}})}]{Cliff:2005}
Grissom RJ, Kim JJ (2005{\natexlab{a}}) Effect sizes for research: A broad
  practical approach, 2nd edn. Lawrence Earlbaum Associates

\bibitem[{Grissom and Kim(2005{\natexlab{b}})}]{cliff}
Grissom RJ, Kim JJ (2005{\natexlab{b}}) Effect sizes for research: A broad
  practical approach, 2nd edn. Lawrence Earlbaum Associates

\bibitem[{Hall et~al(2009)Hall, Frank, Holmes, Pfahringer, Reutemann, and
  Witten}]{Weka}
Hall M, Frank E, Holmes G, Pfahringer B, Reutemann P, Witten IH (2009) The weka
  data mining software: An update. SIGKDD Explor Newsl 11(1):10--18,
  \doi{10.1145/1656274.1656278},
  \urlprefix\url{http://doi.acm.org/10.1145/1656274.1656278}

\bibitem[{Hall et~al(2011)Hall, Beecham, Bowes, Gray, and
  Counsell}]{HallBBGC11}
Hall T, Beecham S, Bowes D, Gray D, Counsell S (2011) Developing
  fault-prediction models: What the research can show industry. {IEEE} Software
  28(6):96--99

\bibitem[{Han et~al(2008)Han, Jeon, Bae, and Hong}]{han2008behavioral}
Han AR, Jeon SU, Bae DH, Hong JE (2008) Behavioral dependency measurement for
  change-proneness prediction in uml 2.0 design models. In: 32nd Annual IEEE
  International Computer Software and Applications Conference, IEEE, pp 76--83

\bibitem[{Han et~al(2010)Han, Jeon, Bae, and Hong}]{han2010measuring}
Han AR, Jeon SU, Bae DH, Hong JE (2010) Measuring behavioral dependency for
  improving change-proneness prediction in uml-based design models. Journal of
  Systems and Software 83(2):222--234

\bibitem[{Hassan(2009)}]{hassan2009predicting}
Hassan AE (2009) Predicting faults using the complexity of code changes. In:
  Int'l Conf. Software Engineering (ICSE), IEEE, pp 78--88

\bibitem[{John and Langley(1995)}]{John:1995}
John GH, Langley P (1995) Estimating continuous distributions in bayesian
  classifiers. In: Eleventh Conference on Uncertainty in Artificial
  Intelligence, Morgan Kaufmann, San Mateo, pp 338--345

\bibitem[{Kennedy(2011)}]{kennedy2011particle}
Kennedy J (2011) Particle swarm optimization. In: Encyclopedia of machine
  learning, Springer, pp 760--766

\bibitem[{Kessentini et~al(2010)Kessentini, Vaucher, and
  Sahraoui}]{Kessentini:ase2010}
Kessentini M, Vaucher S, Sahraoui H (2010) Deviance from perfection is a better
  criterion than closeness to evil when identifying risky code. In: Proceedings
  of the IEEE/ACM International Conference on Automated Software Engineering,
  ACM, ASE '10, pp 113--122

\bibitem[{Kessentini et~al(2014)Kessentini, Kessentini, Sahraoui, Bechikh, and
  Ouni}]{Kessentini-sbse:tse}
Kessentini W, Kessentini M, Sahraoui H, Bechikh S, Ouni A (2014) A cooperative
  parallel search-based software engineering approach for code-smells
  detection. IEEE Transactions on Software Engineering 40(9):841--861,
  \doi{10.1109/TSE.2014.2331057}

\bibitem[{Khomh et~al(2009{\natexlab{a}})Khomh, Di~Penta, and
  Gueheneuc}]{khomh2009exploratory}
Khomh F, Di~Penta M, Gueheneuc YG (2009{\natexlab{a}}) An exploratory study of
  the impact of code smells on software change-proneness. In: 2009 16th Working
  Conference on Reverse Engineering, IEEE, pp 75--84

\bibitem[{Khomh et~al(2009{\natexlab{b}})Khomh, Vaucher, Gu{\'e}h{\'e}neuc, and
  Sahraoui}]{Khomh:qsic2009}
Khomh F, Vaucher S, Gu{\'e}h{\'e}neuc YG, Sahraoui H (2009{\natexlab{b}}) A
  bayesian approach for the detection of code and design smells. In:
  Proceedings of the International Conference on Quality Software (QSIC), IEEE,
  Hong Kong, China, pp 305--314

\bibitem[{Khomh et~al(2012)Khomh, Di~Penta, Gu{\'e}h{\'e}neuc, and
  Antoniol}]{khomh2012exploratory}
Khomh F, Di~Penta M, Gu{\'e}h{\'e}neuc YG, Antoniol G (2012) An exploratory
  study of the impact of antipatterns on class change-and fault-proneness.
  Empirical Softw Engg 17(3):243--275

\bibitem[{Kohavi(1995)}]{Kohavi:1995}
Kohavi R (1995) The power of decision tables. In: 8th European Conference on
  Machine Learning, Springer, pp 174--189

\bibitem[{Kumar et~al(2017{\natexlab{a}})Kumar, Behera, Rath, and
  Sureka}]{kumar2017transfer}
Kumar L, Behera RK, Rath S, Sureka A (2017{\natexlab{a}}) Transfer learning for
  cross-project change-proneness prediction in object-oriented software
  systems: A feasibility analysis. ACM SIGSOFT Software Engineering Notes
  42(3):1--11

\bibitem[{Kumar et~al(2017{\natexlab{b}})Kumar, Rath, and
  Sureka}]{kumar2017empirical}
Kumar L, Rath SK, Sureka A (2017{\natexlab{b}}) Empirical analysis on
  effectiveness of source code metrics for predicting change-proneness. In:
  ISEC, pp 4--14

\bibitem[{Lanza and Marinescu(2006)}]{Lanza:2006}
Lanza M, Marinescu R (2006) Object-Oriented Metrics in Practice: Using Software
  Metrics to Characterize, Evaluate, and Improve the Design of Object-Oriented
  Systems. Springer

\bibitem[{Le~Cessie and Van~Houwelingen(1992)}]{le1992ridge}
Le~Cessie S, Van~Houwelingen JC (1992) Ridge estimators in logistic regression.
  Applied statistics pp 191--201

\bibitem[{Lehman and Belady(1985)}]{lehman:1985}
Lehman MM, Belady LA (eds)  (1985) Program Evolution: Processes of Software
  Change. Academic Press Professional, Inc.

\bibitem[{Lu et~al(2012)Lu, Zhou, Xu, Leung, and Chen}]{lu2012ability}
Lu H, Zhou Y, Xu B, Leung H, Chen L (2012) The ability of object-oriented
  metrics to predict change-proneness: a meta-analysis. Empirical software
  engineering 17(3):200--242

\bibitem[{Malhotra and Bansal(2015)}]{malhotra2015predicting}
Malhotra R, Bansal A (2015) Predicting change using software metrics: A review.
  In: Int'l Conf. on Reliability, Infocom Technologies and Optimization
  (ICRITO), IEEE, pp 1--6

\bibitem[{Malhotra and Khanna(2013)}]{malhotra2013investigation}
Malhotra R, Khanna M (2013) Investigation of relationship between
  object-oriented metrics and change proneness. International Journal of
  Machine Learning and Cybernetics 4(4):273--286

\bibitem[{Malhotra and Khanna(2014)}]{malhotra2014new}
Malhotra R, Khanna M (2014) A new metric for predicting software change using
  gene expression programming. In: Proc. Int'l Workshop on Emerging Trends in
  Software Metrics, ACM, pp 8--14

\bibitem[{Malhotra and Khanna(2017)}]{malhotra2017software}
Malhotra R, Khanna M (2017) Software change prediction using voting particle
  swarm optimization based ensemble classifier. In: Proc. of the Genetic and
  Evolutionary Computation Conference Companion, ACM, pp 311--312

\bibitem[{Marinescu(2014)}]{marinescu2014good}
Marinescu C (2014) How good is genetic programming at predicting changes and
  defects? In: Int'l Symp. on Symbolic and Numeric Algorithms for Scientific
  Computing (SYNASC), IEEE, pp 544--548

\bibitem[{Marinescu(2004)}]{Marinescu04}
Marinescu R (2004) Detection strategies: Metrics-based rules for detecting
  design flaws. In: Proceedings of the International Conference on Software
  Maintenance (ICSM), pp 350--359

\bibitem[{Marinescu(2012)}]{marinescu2012assessing}
Marinescu R (2012) Assessing technical debt by identifying design flaws in
  software systems. IBM Journal of Research and Development 56(5):9--1

\bibitem[{Menzies et~al(2012)Menzies, Caglayan, Kocaguneli, Krall, Peters, and
  Turhan}]{menzies2012promise}
Menzies T, Caglayan B, Kocaguneli E, Krall J, Peters F, Turhan B (2012) The
  promise repository of empirical software engineering data

\bibitem[{Miryung~Kim(2014)}]{kim:tse14}
Miryung~Kim NN Tom~Zimmermann (2014) An empirical study of refactoring
  challenges and benefits at {Microsoft}. IEEE Transactions on Software
  Engineering 40

\bibitem[{Mkaouer et~al(2014)Mkaouer, Kessentini, Bechikh, and
  Cinn{\'e}ide}]{mkaouer2014robust}
Mkaouer MW, Kessentini M, Bechikh S, Cinn{\'e}ide M{\'O} (2014) A robust
  multi-objective approach for software refactoring under uncertainty. In:
  International Symposium on Search Based Software Engineering, Springer, pp
  168--183

\bibitem[{Moha et~al(2010)Moha, Gu{\'e}h{\'e}neuc, Duchien, and
  Meur}]{Moha:tse2010}
Moha N, Gu{\'e}h{\'e}neuc YG, Duchien L, Meur AFL (2010) Decor: A method for
  the specification and detection of code and design smells. IEEE Transactions
  on Software Engineering 36(1):20--36

\bibitem[{Morales et~al(2016)Morales, Soh, Khomh, Antoniol, and
  Chicano}]{Morales:jss16}
Morales R, Soh Z, Khomh F, Antoniol G, Chicano F (2016) On the use of
  developers' context for automatic refactoring of software anti-patterns.
  Journal of Systems and Software (JSS)

\bibitem[{Munro(2005)}]{Munro05-BadSmellIdentification}
Munro MJ (2005) Product metrics for automatic identification of ``bad smell"
  design problems in java source-code. In: Proceedings of the International
  Software Metrics Symposium (METRICS), IEEE, p~15

\bibitem[{Murphy-Hill and Black(2010)}]{murphy2010interactive}
Murphy-Hill E, Black AP (2010) An interactive ambient visualization for code
  smells. In: Proceedings of the 5th international symposium on Software
  visualization, ACM, pp 5--14

\bibitem[{O'brien(2007)}]{Obrien2007}
O'brien RM (2007) A caution regarding rules of thumb for variance inflation
  factors. Quality {\&} Quantity 41(5):673--690

\bibitem[{Olbrich et~al(2010)Olbrich, Cruzes, and Sjøberg}]{Olbrich10areall}
Olbrich SM, Cruzes DS, Sjøberg DIK (2010) Are all code smells harmful? a study
  of god classes and brain classes in the evolution of three open source
  systems. In: in: Int’l Conf. Softw. Maint, pp 1--10

\bibitem[{Oliveto et~al(2010)Oliveto, Khomh, Antoniol, and
  Gu{\'e}h{\'e}neuc}]{Oliveto10-CSMR-ABS}
Oliveto R, Khomh F, Antoniol G, Gu{\'e}h{\'e}neuc YG (2010) Numerical
  signatures of antipatterns: An approach based on {B}-splines. In: Proceedings
  of the European Conference on Software Maintenance and Reengineering (CSMR),
  IEEE, pp 248--251

\bibitem[{Palomba and Zaidman(2017)}]{palomba2017does}
Palomba F, Zaidman A (2017) Does refactoring of test smells induce fixing flaky
  tests? In: Software Maintenance and Evolution (ICSME), 2017 IEEE
  International Conference on, IEEE, pp 1--12

\bibitem[{Palomba et~al(2013)Palomba, Bavota, Di~Penta, Oliveto, De~Lucia, and
  Poshyvanyk}]{palomba2013detecting}
Palomba F, Bavota G, Di~Penta M, Oliveto R, De~Lucia A, Poshyvanyk D (2013)
  Detecting bad smells in source code using change history information. In:
  Automated software engineering (ASE), 2013 IEEE/ACM 28th international
  conference on, IEEE, pp 268--278

\bibitem[{Palomba et~al(2014)Palomba, Bavota, Di~Penta, Oliveto, and
  De~Lucia}]{palomba2014they}
Palomba F, Bavota G, Di~Penta M, Oliveto R, De~Lucia A (2014) Do they really
  smell bad? a study on developers' perception of bad code smells. In: Software
  maintenance and evolution (ICSME), 2014 IEEE international conference on,
  IEEE, pp 101--110

\bibitem[{Palomba et~al(2015{\natexlab{a}})Palomba, Bavota, Di~Penta, Oliveto,
  Poshyvanyk, and De~Lucia}]{Palomba:tse2015}
Palomba F, Bavota G, Di~Penta M, Oliveto R, Poshyvanyk D, De~Lucia A
  (2015{\natexlab{a}}) Mining version histories for detecting code smells. IEEE
  Transactions on Software Engineering 41(5):462--489,
  \doi{10.1109/TSE.2014.2372760}

\bibitem[{Palomba et~al(2015{\natexlab{b}})Palomba, Lucia, Bavota, and
  Oliveto}]{bookchapert:Palomba}
Palomba F, Lucia AD, Bavota G, Oliveto R (2015{\natexlab{b}}) Anti-pattern
  detection: Methods, challenges, and open issues. Advances in Computers
  95:201--238, \doi{10.1016/B978-0-12-800160-8.00004-8}

\bibitem[{Palomba et~al(2016)Palomba, Panichella, Zaidman, Oliveto, and {De
  Lucia}}]{Palomba:icpc2016}
Palomba F, Panichella A, Zaidman A, Oliveto R, {De Lucia} A (2016) A
  textual-based technique for smell detection. In: Proceedings of the 24th
  International Conference on Program Comprehension (ICPC 2016), IEEE, Austin,
  USA, p to appear

\bibitem[{Palomba et~al(2017{\natexlab{a}})Palomba, Bavota, Di~Penta, Fasano,
  Oliveto, and De~Lucia}]{palomba2017diffuseness}
Palomba F, Bavota G, Di~Penta M, Fasano F, Oliveto R, De~Lucia A
  (2017{\natexlab{a}}) On the diffuseness and the impact on maintainability of
  code smells: a large scale empirical investigation. Empirical Software
  Engineering pp 1--34

\bibitem[{Palomba et~al(2017{\natexlab{b}})Palomba, Panichella, Zaidman,
  Oliveto, and De~Lucia}]{palomba2017scent}
Palomba F, Panichella A, Zaidman A, Oliveto R, De~Lucia A (2017{\natexlab{b}})
  The scent of a smell: An extensive comparison between textual and structural
  smells. Transactions on Software Engineering

\bibitem[{Palomba et~al(2017{\natexlab{c}})Palomba, Zanoni, Fontana, De~Lucia,
  and Oliveto}]{palomba2017toward}
Palomba F, Zanoni M, Fontana FA, De~Lucia A, Oliveto R (2017{\natexlab{c}})
  Toward a smell-aware bug prediction model. IEEE Transactions on Software
  Engineering

\bibitem[{Palomba et~al(2018{\natexlab{a}})Palomba, Bavota, Di~Penta, Fasano,
  Oliveto, and De~Lucia}]{palomba2018large}
Palomba F, Bavota G, Di~Penta M, Fasano F, Oliveto R, De~Lucia A
  (2018{\natexlab{a}}) A large-scale empirical study on the lifecycle of code
  smell co-occurrences. Information and Software Technology 99:1--10

\bibitem[{Palomba et~al(2018{\natexlab{b}})Palomba, Zaidman, and {De
  Lucia}}]{Palomba:icsme18}
Palomba F, Zaidman A, {De Lucia} A (2018{\natexlab{b}}) {Automatic Test Smell
  Detection using Information Retrieval Techniques}. In: International
  Conference on Software Maintenance and Evolution (ICSME), IEEE, p to appear

\bibitem[{Parnas(1994)}]{parnas}
Parnas DL (1994) Software aging. In: Proc. of the International Conference on
  Software Engineering (ICSE), IEEE, pp 279--287

\bibitem[{Peer and Malhotra(2013)}]{peer2013application}
Peer A, Malhotra R (2013) Application of adaptive neuro-fuzzy inference system
  for predicting software change proneness. In: Advances in Computing,
  Communications and Informatics (ICACCI), 2013 International Conference on,
  IEEE, pp 2026--2031

\bibitem[{Peng et~al(2002)Peng, Lee, and Ingersoll}]{Chao}
Peng CYJ, Lee KL, Ingersoll GM (2002) An introduction to logistic regression
  analysis and reporting. The Journal of Educational Research 96(1):3--14

\bibitem[{Peters and Zaidman(2012)}]{peters2012evaluating}
Peters R, Zaidman A (2012) Evaluating the lifespan of code smells using
  software repository mining. In: Software Maintenance and Reengineering
  (CSMR), 2012 16th European Conference on, IEEE, pp 411--416

\bibitem[{Quinlan(1986)}]{Quinlan:1986}
Quinlan JR (1986) Induction of decision trees. Mach Learn 1(1):81--106,
  \doi{10.1023/A:1022643204877},
  \urlprefix\url{http://dx.doi.org/10.1023/A:1022643204877}

\bibitem[{Ratiu et~al(2004)Ratiu, Ducasse, G\^{\i}rba, and
  Marinescu}]{RatiuDGM04}
Ratiu D, Ducasse S, G\^{\i}rba T, Marinescu R (2004) Using history information
  to improve design flaws detection. In: Proceedings of the European Conference
  on Software Maintenance and Reengineering (CSMR), IEEE, pp 223--232

\bibitem[{Romano and Pinzger(2011)}]{romano2011using}
Romano D, Pinzger M (2011) Using source code metrics to predict change-prone
  java interfaces. In: Proc. Int'l Conf. Software Maintenance (ICSM), IEEE, pp
  303--312

\bibitem[{Rosenblatt(1961)}]{Rosenblatt:1961}
Rosenblatt F (1961) Principles of Neurodynamics: Perceptrons and the Theory of
  Brain Mechanisms. Spartan Books

\bibitem[{Rumbaugh et~al(2004)Rumbaugh, Jacobson, and Booch}]{rumbaugh:2004}
Rumbaugh J, Jacobson I, Booch G (2004) Unified Modeling Language Reference
  Manual, The (2Nd Edition). Pearson Higher Education

\bibitem[{Sahin et~al(2014)Sahin, Kessentini, Bechikh, and
  Deb}]{Kessentini-bilevel:tosem}
Sahin D, Kessentini M, Bechikh S, Deb K (2014) Code-smell detection as a
  bilevel problem. ACM Trans Softw Eng Methodol 24(1):6:1--6:44,
  \doi{10.1145/2675067}

\bibitem[{Scott and Knott(1974)}]{ScottKnott}
Scott AJ, Knott M (1974) A cluster analysis method for grouping means in the
  analysis of variance. Biometrics 30:507--512

\bibitem[{Sharafat and Tahvildari(2007)}]{sharafat2007probabilistic}
Sharafat AR, Tahvildari L (2007) A probabilistic approach to predict changes in
  object-oriented software systems. In: Proc. Conf. on Softw. Maintenance and
  Reengineering (CSMR), IEEE, pp 27--38

\bibitem[{Sharafat and Tahvildari(2008)}]{sharafat2008change}
Sharafat AR, Tahvildari L (2008) Change prediction in object-oriented software
  systems: A probabilistic approach. Journal of Software 3(5):26--39

\bibitem[{Shepperd et~al(2013)Shepperd, Song, Sun, and Mair}]{shepperd2013data}
Shepperd M, Song Q, Sun Z, Mair C (2013) Data quality: Some comments on the
  nasa software defect datasets. IEEE Transactions on Software Engineering
  39(9):1208--1215

\bibitem[{Shepperd et~al(2014{\natexlab{a}})Shepperd, Bowes, and
  Hall}]{Shepperd:tseResearcherBias}
Shepperd M, Bowes D, Hall T (2014{\natexlab{a}}) Researcher bias: The use of
  machine learning in software defect prediction. IEEE Transactions on Software
  Engineering 40(6):603--616, \doi{10.1109/TSE.2014.2322358}

\bibitem[{Shepperd et~al(2014{\natexlab{b}})Shepperd, Bowes, and
  Hall}]{Shepperd}
Shepperd M, Bowes D, Hall T (2014{\natexlab{b}}) Researcher bias: The use of
  machine learning in software defect prediction. IEEE Trans on Softw
  Engineering 40(6):603--616

\bibitem[{Sjoberg et~al(2013)Sjoberg, Yamashita, Anda, Mockus, and
  Dyba}]{sjoberg2013quantifying}
Sjoberg DI, Yamashita A, Anda BC, Mockus A, Dyba T (2013) Quantifying the
  effect of code smells on maintenance effort. IEEE Transactions on Software
  Engineering (8):1144--1156

\bibitem[{Soetens et~al(2016)Soetens, Demeyer, Zaidman, and
  P{\'e}rez}]{soetens:2016}
Soetens QD, Demeyer S, Zaidman A, P{\'e}rez J (2016) Change-based test
  selection: An empirical evaluation. Empirical Softw Engg 21(5):1990--2032

\bibitem[{Song et~al(2011)Song, Jia, Shepperd, Ying, and Liu}]{Song:tse11}
Song Q, Jia Z, Shepperd M, Ying S, Liu J (2011) A general software
  defect-proneness prediction framework. IEEE Transactions on Software
  Engineering 37(3):356--370, \doi{10.1109/TSE.2010.90}

\bibitem[{Spadini et~al(2018)Spadini, Palomba, Zaidman, Bruntink, and
  Bacchelli}]{Spadini:icsme18}
Spadini D, Palomba F, Zaidman A, Bruntink M, Bacchelli A (2018) {On The
  Relation of Test Smells to Software Code Quality}. In: International
  Conference on Software Maintenance and Evolution (ICSME), IEEE, p to appear

\bibitem[{Spinellis(2005)}]{spinellis2005tool}
Spinellis D (2005) Tool writing: A forgotten art? IEEE Software (4):9--11

\bibitem[{Stone(1974)}]{stone1974cross}
Stone M (1974) Cross-validatory choice and assessment of statistical
  predictions. Journal of the royal statistical society Series B
  (Methodological) pp 111--147

\bibitem[{Taba et~al(2013)Taba, Khomh, Zou, Hassan, and Nagappan}]{Taba:2013}
Taba SES, Khomh F, Zou Y, Hassan AE, Nagappan M (2013) Predicting bugs using
  antipatterns. In: Proceedings of the 2013 IEEE International Conference on
  Software Maintenance, IEEE Computer Society, Washington, DC, USA, ICSM '13,
  pp 270--279, \doi{10.1109/ICSM.2013.38},
  \urlprefix\url{http://dx.doi.org/10.1109/ICSM.2013.38}

\bibitem[{Taibi et~al(2017)Taibi, Janes, and Lenarduzzi}]{taibi2017developers}
Taibi D, Janes A, Lenarduzzi V (2017) How developers perceive smells in source
  code: A replicated study. Information and Software Technology 92:223--235

\bibitem[{Tantithamthavorn et~al(2016{\natexlab{a}})Tantithamthavorn, McIntosh,
  Hassan, and Matsumoto}]{Tantithamthavorn:icse2016}
Tantithamthavorn C, McIntosh S, Hassan AE, Matsumoto K (2016{\natexlab{a}})
  Automated parameter optimization of classification techniques for defect
  prediction models. In: Proceedings of the 38th International Conference on
  Software Engineering, ACM, New York, NY, USA, ICSE '16, pp 321--332,
  \doi{10.1145/2884781.2884857},
  \urlprefix\url{http://doi.acm.org/10.1145/2884781.2884857}

\bibitem[{Tantithamthavorn et~al(2016{\natexlab{b}})Tantithamthavorn, McIntosh,
  Hassan, and Matsumoto}]{Tantithamthavorn:comments}
Tantithamthavorn C, McIntosh S, Hassan AE, Matsumoto K (2016{\natexlab{b}})
  Comments on researcher bias: The use of machine learning in software defect
  prediction. IEEE Transactions on Software Engineering 42(11):1092--1094,
  \doi{10.1109/TSE.2016.2553030}

\bibitem[{Tantithamthavorn et~al(2017)Tantithamthavorn, McIntosh, Hassan, and
  Matsumoto}]{Tantithamthavorn:2017}
Tantithamthavorn C, McIntosh S, Hassan AE, Matsumoto K (2017) An empirical
  comparison of model validation techniques for defect prediction models. IEEE
  Trans Softw Eng 43(1):1--18, \doi{10.1109/TSE.2016.2584050},
  \urlprefix\url{https://doi.org/10.1109/TSE.2016.2584050}

\bibitem[{Tempero et~al(2010)Tempero, Anslow, Dietrich, Han, Li, Lumpe, Melton,
  and Noble}]{Tempero2010}
Tempero E, Anslow C, Dietrich J, Han T, Li J, Lumpe M, Melton H, Noble J (2010)
  The qualitas corpus: A curated collection of java code for empirical studies.
  In: Proc. 17th Asia Pacific Software Eng. Conf., IEEE, Sydney, Australia, pp
  336--345, \doi{10.1109/APSEC.2010.46}

\bibitem[{Theodoridis and Koutroumbas(2008)}]{theodoridis2008pattern}
Theodoridis S, Koutroumbas K (2008) Pattern recognition. IEEE Transactions on
  Neural Networks 19(2):376--376

\bibitem[{Tsantalis and Chatzigeorgiou(2009)}]{Tsantalis:tse2009}
Tsantalis N, Chatzigeorgiou A (2009) Identification of move method refactoring
  opportunities. IEEE Transactions on Software Engineering 35(3):347--367

\bibitem[{Tsantalis et~al(2005)Tsantalis, Chatzigeorgiou, and
  Stephanides}]{tsantalis2005predicting}
Tsantalis N, Chatzigeorgiou A, Stephanides G (2005) Predicting the probability
  of change in object-oriented systems. IEEE Transactions on Software
  Engineering 31(7):601--614

\bibitem[{Tufano et~al(2015)Tufano, Palomba, Bavota, Oliveto, Di~Penta,
  De~Lucia, and Poshyvanyk}]{tufano2015and}
Tufano M, Palomba F, Bavota G, Oliveto R, Di~Penta M, De~Lucia A, Poshyvanyk D
  (2015) When and why your code starts to smell bad. In: Proceedings of the
  37th International Conference on Software Engineering-Volume 1, IEEE Press,
  pp 403--414

\bibitem[{Tufano et~al(2016)Tufano, Palomba, Bavota, Di~Penta, Oliveto,
  De~Lucia, and Poshyvanyk}]{tufano2016empirical}
Tufano M, Palomba F, Bavota G, Di~Penta M, Oliveto R, De~Lucia A, Poshyvanyk D
  (2016) An empirical investigation into the nature of test smells. In:
  Proceedings of the 31st IEEE/ACM International Conference on Automated
  Software Engineering, pp 4--15

\bibitem[{Tufano et~al(2017)Tufano, Palomba, Bavota, Oliveto, Di~Penta,
  De~Lucia, and Poshyvanyk}]{Tufano:icse2015}
Tufano M, Palomba F, Bavota G, Oliveto R, Di~Penta M, De~Lucia A, Poshyvanyk D
  (2017) When and why your code starts to smell bad (and whether the smells go
  away). IEEE Transactions on Software Engineering p to appear.

\bibitem[{Vidal et~al(2016{\natexlab{a}})Vidal, Guimaraes, Oizumi, Garcia,
  Pace, and Marcos}]{vidal2016criteria}
Vidal S, Guimaraes E, Oizumi W, Garcia A, Pace AD, Marcos C
  (2016{\natexlab{a}}) On the criteria for prioritizing code anomalies to
  identify architectural problems. In: Proceedings of the 31st Annual ACM
  Symposium on Applied Computing, ACM, pp 1812--1814

\bibitem[{Vidal et~al(2016{\natexlab{b}})Vidal, Marcos, and
  D{\'\i}az-Pace}]{vidal2016approach}
Vidal SA, Marcos C, D{\'\i}az-Pace JA (2016{\natexlab{b}}) An approach to
  prioritize code smells for refactoring. Automated Software Engineering
  23(3):501--532

\bibitem[{Y.~Freund(1999)}]{Freund:1999}
Y~Freund LM (1999) The alternating decision tree learning algorithm. In:
  Proceeding of the Sixteenth International Conference on Machine Learning, pp
  124--133

\bibitem[{Yamashita and Moonen(2013)}]{Yamashita:ICSE2013}
Yamashita A, Moonen L (2013) Exploring the impact of inter-smell relations on
  software maintainability: An empirical study. In: Proceedings of the
  International Conference on Software Engineering (ICSE), IEEE, pp 682--691

\bibitem[{Yamashita and Moonen(2012)}]{YamashitaM12}
Yamashita AF, Moonen L (2012) Do code smells reflect important maintainability
  aspects? In: Proceedings of the International Conference on Software
  Maintenance (ICSM), IEEE, pp 306--315

\bibitem[{Zhao and Hayes(2011)}]{zhao2011rank}
Zhao L, Hayes JH (2011) Rank-based refactoring decision support: two studies.
  Innovations in Systems and Software Engineering 7(3):171

\bibitem[{Zhou et~al(2009)Zhou, Leung, and Xu}]{zhou2009examining}
Zhou Y, Leung H, Xu B (2009) Examining the potentially confounding effect of
  class size on the associations between object-oriented metrics and
  change-proneness. IEEE Transactions on Software Engineering 35(5):607--623

\end{thebibliography}

\end{document}